\RequirePackage{fix-cm}
\documentclass[smallextended]{svjour3}       
\smartqed  
\usepackage{graphicx}
\usepackage{wrapfig}
\usepackage{geometry}
\usepackage{amsmath}
\usepackage{multirow,tabularx}
\usepackage{footnote, url}
\usepackage[flushleft]{threeparttable}
\usepackage{mdframed}
\graphicspath{{}{images/}{dia/}}
\DeclareGraphicsExtensions{.pdf,.png}
\newcounter{o}
\usepackage{xspace}
\usepackage{tikz}
\usepackage{framed}
\usepackage{adjustbox}
\usepackage{scalefnt}
\usepackage{pgfplots, pgfplotstable}
\usepackage[bookmarks, bookmarksopen=true, plainpages=false, pdfpagelabels, pdfpagelayout=SinglePage, breaklinks=true]{hyperref}
\pgfplotsset{compat=1.16}
\usepackage{hyperref}
\usepackage{listings}
\usepackage{longtable,tabu}
\usepackage[font=small]{caption}
\pgfplotsset{compat=1.16}
\usepackage{soul}

\usepackage{parcolumns}

\usepackage{subcaption}
\usepackage{amsfonts}

\usepackage{xcolor}
\usepackage{tcolorbox}
\tcbuselibrary{skins, breakable, theorems}

\newcommand{\edit}[1]{\textcolor{black}{#1 }}
\newcommand{\minor}[1]{\textcolor{black}{#1 }}

\definecolor{Gray}{gray}{0.9}

\definecolor{codegreen}{rgb}{0,0.6,0}
\definecolor{codegray}{rgb}{0.5,0.5,0.5}
\definecolor{codegray2}{rgb}{242, 243, 244}
\definecolor{codepurple}{rgb}{0.58,0,0.82}
\definecolor{backcolour}{rgb}{189, 195, 199}

\lstdefinestyle{mystyle}{
  backgroundcolor=\color{gray!10}, 
  commentstyle=\sffamily\raggedright\color{codegreen},
  keywordstyle=\color{blue}\bf,
  numberstyle=\tiny\color{black},
  stringstyle=\color{codepurple},
  basicstyle=\ttfamily\tiny,
  breakatwhitespace=false,         
  breaklines=true,                 
  captionpos=b,                    
  keepspaces=true,                 
  numbers=left, 
  stepnumber=1,
  xleftmargin=3.4pt,
  xrightmargin=3.4pt,
  numbersep=5pt,                  
  showspaces=false,                
  showstringspaces=false,
  showtabs=false,                  
  tabsize=2,
  language=Python,
  frame=single
}

\usepackage{enumitem}
\setlist[itemize]{noitemsep, topsep=0pt}

\newcolumntype{Y}{>{\centering\arraybackslash}X}

\newcolumntype{g}{>{\columncolor{Gray}}c}
\newcolumntype{n}{>{\columncolor{Gray}}r}

\newcolumntype{v}{>{\columncolor{Gray}}m{0.75cm}}
\newcolumntype{y}{>{\columncolor{Gray}}m{1.6cm}}

\newcolumntype{x}{>{\columncolor{Gray}}m{1.7cm}}

\newcolumntype{z}{>{\columncolor{Gray}}m{3.85cm}}

\newcolumntype{k}{>{\columncolor{codegray2}}l}

\definecolor{codegray}{rgb}{0.95,0.95,0.92}
\sethlcolor{codegray}

\tcbset{
bgtable/.style={
enhanced,
colframe=red!50!black,
colbacktitle=Salmon!30,
coltitle=black,
center title,
fonttitle=\bfseries\sffamily,
watermark graphics=#1,
watermark opacity=0.3,
nobeforeafter}
}

\definecolor{fbtitle}{HTML}{636463}
\definecolor{fbbg}{HTML}{F2F2F2}

\newtcbtheorem{answer}{Findings}%
  {enhanced, breakable,
    colback = fbbg, colframe = gray, colbacktitle = fbtitle,
    attach boxed title to top left = {yshift = -2.2mm, xshift = 0mm},
    boxed title style = {rounded corners},
    separator sign none,
    fonttitle = \sffamily\bfseries}{qst}

\usepackage{footmisc}

\usepackage[compact]{titlesec}
\newcommand{\rqone}{\textbf{RQ1:} \textit{How does altering code at various granularity affect the model's ability to generate functionally correct code? }}

\newcommand{\rqtwo}{\textbf{RQ2:} \textit{How do modifications in natural language descriptions at different granularity levels influence the functional accuracy of the code generated by the model?}}

\newcommand{\rqthree}{\textbf{RQ3:} \textit{How does changing comments within the code at various levels impact the accuracy and functionality of the code generated by the model?}}

\journalname{Empirical Software Engineering}

\usepackage{pifont}
\usepackage{tikz}
\usetikzlibrary{fadings}
\usetikzlibrary{shapes}
\usetikzlibrary{shadows.blur}
\usepackage{etoolbox}
\newcommand{\circled}[2][]{%
  \tikz[baseline=(char.base)]{%
    \node[shape=circle,
    draw, 
    color=black!30!blue,
    fill=gray!20,
    inner sep=0.5pt]
    (char) {\phantom{\ifblank{#1}{#2}{#1}}};%
    \node at (char.center) {\color{blue}\bfseries\sffamily #2};}}
\robustify{\circled}
\newcommand{\dcircled}[1]{\circled[0]{#1}}

\usepackage{circledtext}
\usepackage{pgffor}
\usepackage{xstring}
\usepackage{fancyvrb}

\usepackage{xurl}

\begin{document}

\title{Adversarial Attack Classification and Robustness Testing for Large Language Models for Code}

\titlerunning{Adversarial Attack Classification and Robustness Testing for Large Language Models for Code}

\author{Yang Liu \and Armstrong Foundjem \and  Foutse Khomh \and  Heng Li
}


\institute{Yang Liu; Armstrong Foundjem; Foutse Khomh; Heng Li \at
              Polytechnique Montréal \\
              \email{\{Yang-2.liu,\;a.foundjem,\;foutse.khomh,\;heng.li\}@polymtl.ca}           
}

\date{Received: date / Accepted: date}

\maketitle

\begin{abstract}

\noindent\textbf{Context:}
In the rapidly evolving landscape of software development, \edit{Large Language models have become essential tools for code tasks such as code generation, completion, analysis, and suggestion. The growing integration of Large Language models for Code into development workflows highlights the critical need for ensuring robustness against potential vulnerabilities—specifically, weaknesses in how these models handle various inputs that could lead to incorrect or insecure code completion.} In this context, vulnerabilities can manifest as the model’s susceptibility to producing flawed code when faced with adversarial inputs or subtle perturbations in task descriptions, code, or comments.

\noindent\textbf{Objective:}
Existing works have neglected the role of natural language inputs that guide code tasks. This study explores the impact of natural language, such as prompts, on adversarial attacks against Large Language models for Code. It assesses how perturbations at the level of character, word, and sentence affect model output. Identifying the most harmful vulnerabilities will inform the development of more resilient code-generation tools.

\noindent\textbf{Method:}
We analyzed several key projects (including ReCode and Openattack) and datasets (including HumanEval and MBPP), to systematically examine how different types of attacks affect model performance. The first dimension of our taxonomy categorizes adversarial attacks based on the content type—whether the attack targets the code, natural language prompts, or comments. The second dimension classifies these attacks by their granularity, such as character-level, word-level, or sentence/statement-level perturbations. We employed a mixed-methods approach, combining quantitative analysis of model performance metrics with qualitative assessments to identify specific vulnerabilities across different types of inputs.

\noindent\textbf{Results:} 
Our analysis revealed that \edit{Large Language models for Code (LLM4Code)}exhibit varying levels of robustness depending on the type and granularity of adversarial perturbations. Sentence-level perturbations were generally met with the highest resilience, indicating that \edit{LLM4Code}can maintain code integrity when faced with broader contextual disruptions. However, word-level perturbations posed significant challenges, revealing vulnerabilities in the models' ability to handle disruptions at this semantic level. Character-level perturbations showed mixed results, with models displaying both strengths and weaknesses in addressing minor syntactical deviations. These findings underscore the need for continued advancements in model robustness, particularly against semantic perturbations, to ensure reliable performance in diverse adversarial scenarios.


\noindent\textbf{Conclusions:}
This study presents a comprehensive framework for evaluating the robustness of Large Language models for Code. It highlights the significance of incorporating code and natural language inputs in adversarial testing, which is essential for improving security and reliability in real-world applications.

\end{abstract}

\section{Introduction}\label{sec:intro}
Large Language models for Code (LLM4Code)~\cite{nijkamp2022codegen,fried2022incoder} have become integral tools in software development, offering automated solutions for generating, analyzing, and suggesting code. This advancement represents more than just efficiency; it marks a significant shift in software development practices~\cite{batouta2016automation}. However, the growing reliance on LLM4Code has exposed these models to various adversarial attacks, which can lead to the generation of flawed or insecure code. Ensuring the robustness of LLM4Code against such attacks is critical for maintaining the integrity and security of software systems~\cite{bielik2020adversarial}. These models, acting as co-developers, assist in complex problem-solving and sometimes autonomously generate functional code~\cite{nijkamp2022codegen,fried2022incoder}. This deep integration into the software lifecycle makes them attractive targets for adversarial attacks, where crafted inputs can induce the generation of incorrect or malicious code~\cite{carlini2017towards}. Perturbations—small, intentional modifications to input data—can exploit model vulnerabilities, leading to flawed outputs~\cite{metzen2017detecting}.

\begin{figure*}[!ht]
    \centering
        \begin{subfigure}[t]{0.5\textwidth}
        \centering
        \fbox{
        \includegraphics[scale=0.35]{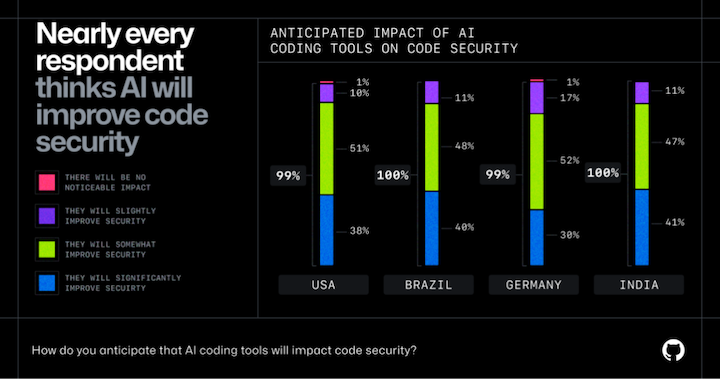}
        }
        \caption{\textbf{GitHub}\textsuperscript{\textcopyright} surveyed 2K software developers, CC BY-NC-SA 2.0}
        \label{fig:sub-a}
    \end{subfigure}
    ~ 
    \begin{subfigure}[t]{0.5\textwidth}
        \centering
        \fbox{\includegraphics[scale=0.17]{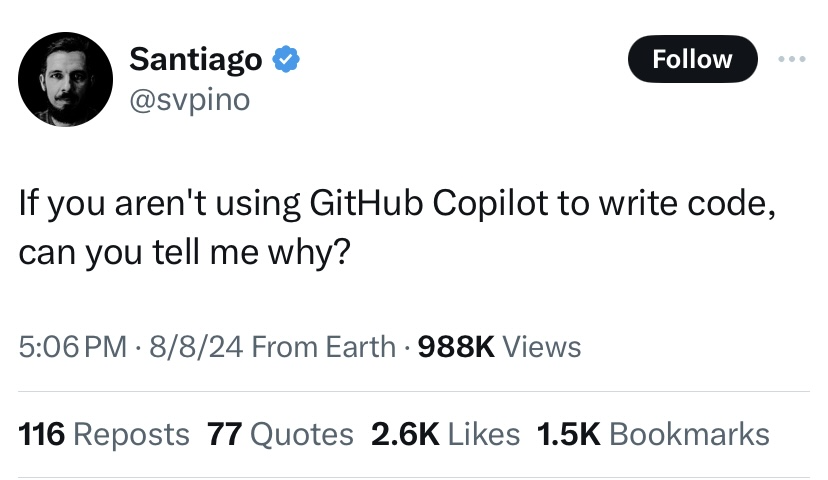}}
    \caption{\textbf{\textit{Santiago}}\textsuperscript{\textcopyright} surveyed 1K software developers, CC BY-NC-SA 2.0}\label{fig:sub-b}
    \end{subfigure}
    \caption{\textbf{\textit{GitHub}} survey (Fig.\ref{fig:sub-a}) indicates 97\% of developers adopting AI tools, highlighting the need for trust and clear guidelines to maximize their effectiveness in software development. Meanwhile, a prominent ML practitioner \textbf{\textit{Santiago Valdarrama}} (Fig.\ref{fig:sub-b}) cautions practitioners: "AI isn't the threat; those who use AI are."}

    \label{fig:dis}
\end{figure*}

The motivation for this study arises from the inherent similarities between programming languages and natural language, both governed by structured grammatical and semantic rules~\cite{ernst2017natural}. LLM4Code serve as a bridge, translating natural language descriptions into executable programs. Given this intersection, it is crucial to explore how perturbations in both the natural language and code components impact LLM4Code performance and output. Survey studies 
by GitHub\footnote{\label{fn1}\url{https://github.blog/news-insights/research/survey-ai-wave-grows/}}, and prominent ML practitioners\footnote{\label{fn2}\url{https://www.linkedin.com/feed/update/urn:li:activity:7228399090647076864/}} highlight the growing adoption of AI-powered tools in software development. For example, the GitHub survey, see Fig.~\ref{fig:sub-a} found that 97\% of developers are incorporating AI tools into their workflows, reflecting their increasing importance. However, this trend also reveals concerns about the reliability and security of these tools. A separate survey by Santiago Valderrama, see Fig.~\ref{fig:sub-b}, conducted on social media, showed that many developers remain skeptical, citing issues like `subtle bugs' introduced by LLM4Code and the need for extensive manual adjustments. The conclusion of both studies suggests that while AI-driven tools like LLM4Code are becoming indispensable in modern software development, 
there is a growing recognition of the robustness and adversarial attack problems\footref{fn1}\footref{fn2} accompanying their use~\cite{anand2021adversarial}. This recognition forms the core motivation for our work: to thoroughly investigate and address these critical issues to enhance the reliability and security of Large Language models for Code in real-world applications. 

Prior research on adversarial attacks has generally focused on either natural language processing (NLP) models or Large Language models for Code, often overlooking the unique challenges 
\edit{such as Strict Syntactical and Semantic, and Context-Specific Logic and Dependencies~\cite{wang2022recode}. Unlike NLP models, where errors may degrade readability but often preserve meaning, even minor perturbations in code (e.g., changing indentation, renaming variables inconsistently) can introduce critical functional errors, making robustness evaluation for LLM4Code particularly challenging.} This study addresses these gaps by introducing a two-dimensional classification system for adversarial attacks on LLM4Code, categorizing them by content type (code, text, comment) and perturbation granularity (character, word, sentence). We also provide a comprehensive evaluation of how these attacks affect LLM4Code robustness, offering deeper insights into potential vulnerabilities and enhancing the understanding of their implications for model security.

\edit{Our work builds upon ReCode~\cite{wang2022recode} and OpenAttack~\cite{zeng2020openattack} but introduces significant new contributions. We expand the adversarial attack classification system used in natural language tasks to code completion, categorizing perturbations based on both perturbation type (character, word, and sentence level) and perturbation target (e.g., identifiers, literals, keywords, structure). Additionally, we incorporate six new perturbations adapted from OpenAttack, including multi-level (character + word level) perturbations, to deepen our analysis of LLM4Code robustness. We also introduce Llama3 and CodeLlama into our evaluation, examining their robustness compared to prior LLM4Code. Surprisingly, our findings indicate that Llama3 (a general model) outperforms CodeLlama (a code-specific model) in certain robustness scenarios, an insight not previously explored in ReCode.}

\edit{Our work is focusing on evaluating the robustness of LLM4Code (Large Language for Code). LLM4Code is a general term for large language models that handle code tasks. LLM4Code could generate code from task description in natural language and complete unfinished code, even though there is only code guide (module import and function definition) and task description. In our work, our dataset distribution reflects this: 62\% of inputs contain only task descriptions with module import and function definition, while 38\% contain both task descriptions and partial code. Both of them are defined as code completion tasks. In summary, we evaluate the robustness of LLM4Code in code completion tasks.}

\edit{This study primarily focuses on code completion from natural language descriptions, where models generate full function implementations based on textual prompts, module import and function definition. Additionally, a subset of tasks involves code completion, where models generate missing portions of a function given partial code. Our dataset distribution reflects this: 62\% of inputs contain only task descriptions with module import and function definition, while 38\% contain both task descriptions and partial code.}

Our methodology involves controlled experiments where LLM4Code~\cite{nijkamp2022codegen,fried2022incoder} are subjected to various perturbations~\cite{wang2022recode,zeng2020openattack}, simulating real-world adversarial scenarios. 

\noindent This study aims to answer the following research questions:
\begin{itemize}
\item \rqone 

\item \rqtwo

\item \rqthree
\end{itemize}

\noindent The findings reveal that LLM4Code exhibit varying degrees of vulnerability to different levels of perturbation and content types, underscoring the need for targeted defense mechanisms. Notably, mono-language models demonstrated superior resilience in certain contexts compared to multi-language models, suggesting potential benefits from specialization.

\edit{The key contributions of this paper are as follows:
\begin{itemize}
    \item [1)] Introduction of a novel classification system for adversarial perturbations in code completion.
    \item [2)] Expansion of adversarial attack types, incorporating six new perturbation techniques from NLP adversarial frameworks.
    \item [3)] Comprehensive evaluation of LLM4Code vulnerabilities, including newer state-of-the-art models (Llama3, CodeLlama).
    \item [4)] Insights into the resilience of general-purpose vs. code-specific LLMs, revealing that general-purpose LLMs can outperform specialized LLM4Code in certain robustness scenarios.
    \item [5)] Extending NLP robustness research into code completion, validating that adversarial perturbation classifications from NLP also apply to programming languages.
\end{itemize}
}
\noindent\textbf{Paper Organization:}  
Section~\ref{sec:background} provides the background, Section~\ref{sec:related_work} reviews related works, Section~\ref{sec:methodolog} describes our methodology, Section~\ref{sec:eval} presents our results, Section~\ref{sec:discussion} discusses our work,  Section~\ref{sec:implication} discusses the implications, Section~\ref{sec:limi} outlines limitations and future work, Section~\ref{sec:threats} examines potential threats to validity, and Section~\ref{sec:conclusions} concludes the paper with a summary of key contributions. 

\section{Background} \label{sec:background}

In this section, we discuss the background of our study. First, we introduce the concept of adversarial inputs and robustness testing for Large Language models for Code. Then, we propose a detailed thread model for adversarial attacks against Large Language models for Code.

\subsection{Adversarial Inputs and Robustness Testing}

\noindent\textbf{Adversarial inputs.} Adversarial inputs on Large Language models for Code~\cite{nijkamp2022codegen}~\cite{fried2022incoder} is an emerging challenge in the field of AI-driven software development. These models, which facilitate the automation of code writing and optimization, are prone to encountering inputs that, while superficially similar, may contain subtle variations intended to disrupt their output. For instance, consider two requests given to a code generation model: ``Generate SQL query for selecting records'' and ``Craft a database query to fetch entries.'' While the intent behind both commands is ostensibly the same, differences in terminology, structure, and specificity can lead to significantly different code outputs. 
The accuracy of input processed by Large Language models for Code varies significantly, influenced by the user's familiarity with programming concepts, the level of detail provided, and the clarity of the request~\cite{chen2021evaluating}. A study reveals that less experienced developers tend to introduce more ambiguity into their inputs, which can mislead models into producing less accurate or incorrect code~\cite{jia2017adversarial}. This highlights a critical challenge in the field of code generation: the need to address adversarial inputs effectively.

\noindent\textbf{Robustness testing.}
To tackle the challenge of adversarial inputs in code generation, robustness testing is essential. This involves generating test inputs that cover various types and granularities of adversarial attacks, thereby allowing for a thorough evaluation of model resilience~\cite{nijkamp2022codegen}~\cite{fried2022incoder}. Understanding the types of perturbations and their impact is crucial, motivating the classification of these adversarial inputs and informing the research questions that guide model improvements. 
\edit{Our adversarial attack classification system can be practically integrated into code completion pipelines at two key stages: (1) Training, where adversarial training techniques augment datasets with perturbed examples to improve model robustness, and (2) Deployment, where real-time input validation detects and mitigates ambiguous inputs before they influence model-generated code. This approach enhances the security, accuracy, and reliability of Large Language models for Code in real-world applications.
}

\subsection{A Threat Model for Adversarial Attacks Against Large Language models for Code}\label{sec:threat-model}
 \edit{Based on the threat model proposed by Jha et al.~\cite{jha2023codeattack}, we analyze the vulnerabilities of Large Language models for Code (LLM4Code) deployed on company-owned or public servers. In this scenario, attackers can exploit \textbf{input prompts} to manipulate LLM4Code behavior, introducing perturbations that degrade model robustness and reliability. These adversarial attacks target \textbf{the operational phase}, affecting real-time code suggestions, automated software development, and AI-assisted coding workflows.
\textbf{Adversary’s Capabilities:}  
The adversary is assumed to have the ability to \emph{manipulate input prompts} during inference. This attack vector can be exploited in two ways: \underline{1. Server-Side Attack} – \textit{The attacker compromises the hosting infrastructure, injecting modified prompts into LLM4Code queries}. \underline{2. Insider Threat} – \textit{A privileged user modifies inputs before submission, influencing LLM4Code outputs maliciously}~\cite{rizvi2024analyzing}.  
The adversary's modifications can range from \emph{character-level perturbations} (e.g., typos, encoding manipulations) to \emph{semantic alterations} (e.g., misleading comments, function renaming, logic distortions).
\textbf{Adversary’s Knowledge:}  
Operating in a black-box setting, the adversary lacks direct access to the LLM4Code's parameters, training data, or internal architecture. Instead, they rely on \emph{input-output interactions} to iteratively refine attack strategies. This constraint mimics real-world adversarial settings, where \emph{automated code assistants (e.g., Copilot, CodeLlama) are queried via API interfaces} without exposing internal logic.
\textbf{Adversary’s Goals:}  
The adversary aims to exploit LLM4Code weaknesses to degrade output quality, causing:  
- \emph{Functional errors} – Code suggestions that introduce logical bugs.  
- \emph{Security vulnerabilities} – Generated code that contains exploitable flaws (e.g., SQL injection, buffer overflow).  
- \emph{Bias introduction} – Adversarial prompts that mislead LLM4Code into generating unfair or biased responses.  
- \emph{Denial of Service (DoS) Effects} – Inputs designed to induce model failures or infinite loops.
\textbf{Mitigation Strategies:}  
To counteract adversarial threats, we propose the following \emph{defensive measures}:
\underline{1. Detection \& Prevention}  
   - \emph{Enhanced Input Validation:} Implement adversarial filtering techniques to detect perturbed inputs before model execution.  
   - \emph{Syntax-Aware Error Handling:} Train models to reject minor adversarial distortions while maintaining flexibility for natural variations.  
\underline{2. Model Hardening}  
   - \emph{Robust Fine-Tuning:} Extend LLM4Code training with adversarial perturbations to increase resilience against input modifications.  
   - \emph{Diversity in Training Data:} Introduce syntactically and semantically perturbed examples in datasets to strengthen generalization.  
\underline{3. Runtime Safeguards}  
   - \emph{Dynamic Code Analysis:} Use security checkers to validate generated outputs before deployment.  
   - \emph{Comment \& Documentation Validation:} Ensure adversarially injected comments do not influence model-generated code behavior.  
   - \emph{Automated Robustness Testing:} Incorporate systematic perturbation testing to detect vulnerabilities proactively.
By implementing these \emph{multi-layered defenses}, we strengthen LLM4Code against adversarial inputs, reducing the likelihood of \emph{incorrect, insecure, or biased outputs}.
}


Based on the threat model proposed by Jha et al.~\cite{jha2023codeattack}, we consider a situation where Large Language models for Code are utilized within servers—either company-owned or public servers. During this usage, the input prompts or direct inputs critical to the operational phase of these models might be susceptible to perturbations. Specifically, these inputs could be compromised by hackers who target the servers, manipulating the prompts to alter the behavior or output of the Large Language models for Code. This scenario allows the adversary to introduce controlled perturbations, enabling an evaluation of the robustness and resilience of the LLM4Code under adversarial conditions.\\
\textbf{Adversary’s Capabilities}: The adversary in our scenario is assumed to have the capability to manipulate the inputs fed into the Large Language models for Code (LLM4Code)
. This manipulation could occur through compromising the server infrastructure where the LLM4Code are hosted, thereby gaining control over the input pipeline. Similarly, a malicious insider with legitimate access to the system could exploit their position to alter input prompts or inject malicious content into the inputs~\cite{rizvi2024analyzing}. Once access is obtained, the adversary can alter input prompts or inject malicious content into the inputs
during the evaluation or operational phases. The adversary's manipulations can range from subtle perturbations at the character, word, or sentence levels to more sophisticated modifications targeting different content types, such as code, task descriptions, or comments.\\
\textbf{Adversary’s Knowledge}: Operating under a black-box scenario, the adversary does not have access to the internal workings of the LLM4Code, including their parameters, architecture, or training data. Instead, the adversary can only interact with the models by submitting inputs and observing the outputs. This setup reflects realistic attack scenarios where the adversary relies on the outputs to iteratively refine their attack strategy without direct access to the underlying systems.\\
\textbf{Adversary’s Goal}: The primary objective of the adversary is to exploit vulnerabilities within the LLM4Code to cause the models to generate incorrect, insecure, or sub-optimal code. By introducing minimal yet impactful perturbations, the adversary seeks to degrade the quality of the generated code, potentially leading to functional errors, security vulnerabilities, or operational failures in the software.\\
\textbf{Mitigation Strategies:} In our research, we focus on identifying potential vulnerabilities in Large Language models for Code by conducting a thorough threat analysis, which examines how adversaries could exploit these models. We employ code inspection and walkthrough~\cite{fronza2020code} to assess these vulnerabilities. After assessing, we recommend several mitigation strategies. These include enhancing input validation processes to neutralize high-impact perturbations, adjusting model sensitivities—particularly for models like CodeLlama that are more susceptible to specific changes—and extending training datasets to include these challenging scenarios. Furthermore, integrating dynamic code analysis tools can help detect and address code anomalies early, while establishing protocols to manage and validate comments will ensure they do not adversely affect model performance.
Moreover, incorporating robustness testing as a mitigation strategy is crucial, as it systematically evaluates the models' resistance to adversarial attacks, allowing for the identification and reinforcement of weak points in the code generation process.

\section{Related Works}  \label{sec:related_work}

The robustness of Large Language models for Code against adversarial attacks~\cite{bielik2020adversarial} is a growing area of research, drawing parallels from various fields including natural language processing and software engineering. In this section, we briefly introduce previous works that are closely related to our study, which includes pre-trained Large Language models for Code, adversarial attacks against natural language models, and adversarial attacks against programming language models (i.e., code models).

\subsection{Pre-trained Large Language models for Code}

In the realm of software development, pre-trained models have significantly advanced the field of code understanding and generation. Our exploration focuses on three principal model types: First, \textit{Code Embeddings}. These models, such as Code2Vec~\cite{alon2019code2vec} and Code2Seq~\cite{alon2018code2seq}, convert code into vector representations, enhancing tasks such as method name prediction from code snippets. Second, \textit{BERT-based Models}. Notably, CodeBERT~\cite{feng2020codebert} and GraphCodeBERT~\cite{guo2020graphcodebert} have been trained on mixed corpora of code and natural language comments, excelling in tasks such as code synthesis and bug fixing. Third, \textit{GPT-based Models} Including CodeGen~\cite{nijkamp2022codegen}, InCoder~\cite{fried2022incoder}, GPT-J~\cite{de2022zero}, CodeLlama~\cite{roziere2023code}, and Llama-3~\cite{huang2024good}, these models are trained on vast datasets and are adept at various coding tasks due to their understanding of both code and natural language. These models stand out for their ability to automate complex programming tasks effectively, representing a leap in productivity and efficiency in software development. 
In our research, in order to evaluate the state-of-the-art Large Language models for Code by different types of adversarial attacks, we choose GPT-based Models as test models, including the CodeGen~\cite{nijkamp2022codegen} family models, the InCoder~\cite{fried2022incoder} family models, GPT-J~\cite{de2022zero}, CodeLlama~\cite{roziere2023code}, and Llama-3~\cite{huang2024good}. 

\subsection{Adversarial Attacks Against Natural Language Models}

Adversarial attacks are specialized inputs designed to confuse machine learning models such as deep neural networks, including large langauge models, causing them to misclassify given inputs. Yuan et al. explored this phenomenon, focusing on its applications, countermeasures, and identifying key challenges and solutions in the field~\cite{yuan2019adversarial}. 
\edit{In terms of text-specific adversarial strategies}, BAE (BERT-based Adversarial Examples)~\cite{garg2020bae} 
generates adversarial examples by manipulating token placements using BERT’s masked language model capabilities. Another innovative approach is ANTHRO, introduced by Le et al., which leverages real-world text perturbations to conduct more realistic adversarial attacks, showing notable effectiveness against models such as BERT and RoBERTa~\cite{le2022perturbations}. In our study, we utilize the OpenAttack toolkit, an open-source toolkit that provides a comprehensive suite of 15 adversarial attack models, noted for its flexibility and extensibility~\cite{szegedy2013intriguing}, to introduce perturbations within the task description sections of our selected datasets. This approach allows us to systematically assess the resilience of Large Language models for Code to textual modifications in their inputs.

\subsection{Adversarial Attacks Against Code Models}

There are some preivious work on adversarial attacks in Machine Learning for code. Wang et al.~\cite{wang2022recode} propose some perturbations on code and evaluate the robustness of Large Language models for Code. Jha et al.~\cite{jha2023codeattack} propose CodeAttack which is a robustness black-box adversarial attack model that leverages the inherent structure of code to craft subtle yet effective adversarial samples. It efficiently exposes vulnerabilities within cutting-edge programming language (PL) models by demonstrating their susceptibility to code-specific adversarial attacks. 

However, these existing studies have certain limitations. First, adversarial attacks are not systematically categorized, which restricts a more granular understanding of the impact of different perturbations. Second, these works largely focus on attacks targeting only the code itself, neglecting natural language inputs such as task descriptions or comments, which are essential components in the functionality of Large Language models for Code. Third, the number of evaluated models tends to be relatively small, limiting the generalizability of the findings.

To address these gaps, our study~\cite{replication} introduces a novel classification system for analyzing adversarial attacks on Large Language models for Code, categorizing them across two dimensions: granularity and content type. The granularity dimension is segmented into character, word, and sentence/statement levels, allowing for detailed analysis of perturbations at varying depths. The content type dimension categorizes perturbations based on their target within the Large Language models for Code, specifically identifying code, text, and comments. This dual-dimensional framework enables a comprehensive evaluation of vulnerabilities and their potential impacts on the robustness of Large Language models for Code.

\edit{\subsection{Robustness in LLM4Code: Recent Advances and Our Contribution}
Recent works have systematically explore the robustness of code generation models under real-world or adversarial conditions. Mastropaolo et al.~\cite{mastropaolo2023robustness} conducted an empirical evaluation of GitHub Copilot’s resilience to faulty and incomplete prompts, revealing how subtle input changes can significantly degrade output quality. Their study primarily focuses on evaluating a single commercial tool using human-authored prompts and manually crafted perturbations. In contrast, Improta et al.~\cite{improta2025enhancing} investigated strategies for enhancing robustness in AI-based offensive code generators via data augmentation. Their approach emphasizes improving model output quality through training-time interventions, particularly for sensitive or security-critical code. While both studies underscore the importance of robustness, our work complements and extends these efforts by introducing a systematic perturbation classification framework and conducting a granular robustness evaluation across multiple LLM4Code models, input types (code, task description, comments), and perturbation levels (character, word, sentence). Unlike~\cite{mastropaolo2023robustness}, which focuses on a single black-box system, our work evaluates open-source models, and unlike~\cite{improta2025enhancing}, we emphasize evaluation rather than enhancement through retraining. Moreover, our robustness metrics (RPs@k, RDs@k, RRs@k) provide a novel lens for quantifying model degradation under attack. Together, these studies reinforce the critical need for robustness assessment and contribute complementary insights to the emerging discourse on trustworthy LLMs for Code.
}

\section{Methodology} \label{sec:methodolog}
\edit{
To evaluate the robustness of Large Language models for Code (LLM4Code), we categorize adversarial perturbations based on both granularity (character, word, sentence) and target location (task descriptions, comments, code elements). This classification provides a structured framework for assessing how different perturbation types influence model outputs. In this section, we detail our methodology for defining and organizing these perturbations, ensuring that our robustness evaluation is both systematic and reproducible. This analysis is essential for addressing RQ1, RQ2, and RQ3, as it establishes the basis for testing LLM4Code performance under controlled adversarial conditions.}
\subsection{Overview} \label{sec:overview}

\begin{figure*}[ht]
    \centering
    \includegraphics[scale=0.11]{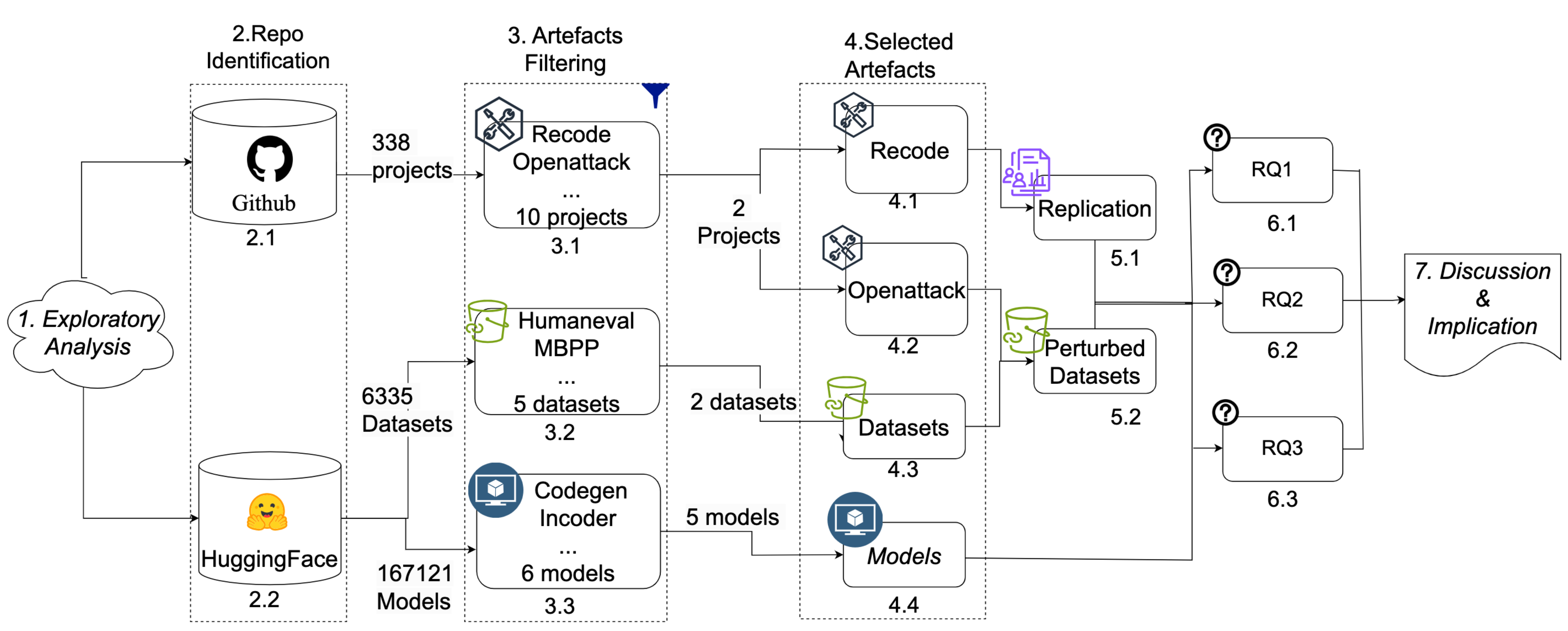}
    \caption{Our proposed approach in this study to address our RQs}
    \label{fig:timeline}
\end{figure*}

\edit{As shown in Fig.~\ref{fig:timeline}, initially, we conducted an exploratory analysis across well-known platforms, including GitHub and Hugging Face, to identify resources relevant to robustness evaluation in Large Language models for Code. On GitHub, we searched for projects using keywords such as ``Adversarial Attacks for Code,'' ``Adversarial Attacks for NLP'', and ``Robustness for Code completion'', retrieving 338 projects relevant to perturbation techniques and LLM4Code robustness evaluation.
For Hugging Face, we found that keyword-based search was ineffective due to the platform’s strict exact-match search functionality, which returned limited or no results for general research topics. Instead, we leveraged Hugging Face’s built-in task selection filters, specifically selecting ``Text Generation'' and ``Text2Text Generation'', which resulted in 157,121 models and 6,335 datasets. We then applied the following structured selection criteria:
\dcircled{1} \textbf{Availability -} Prioritized publicly accessible datasets with no restrictive licenses or missing data.
\dcircled{2} \textbf{Compatibility -} Ensured datasets were suitable for Python-based experimental frameworks such as OpenAttack and ReCode.
\dcircled{3} \textbf{Relevance -} Focused on datasets explicitly related to code completion and robustness evaluation, filtering out purely natural language generation datasets.
\dcircled{4} \textbf{Benchmark Alignment -} Prioritized widely used robustness testing datasets, such as HumanEval and MBPP, which align closely with the objectives of this study.
Hence, the task-driven selection strategy allowed us to comprehensively identify and analyze relevant datasets and models, overcoming the limitations of strict keyword-based searches on Hugging Face.”
}

\begin{figure*}[ht]
    \centering
    \includegraphics[scale=0.3]{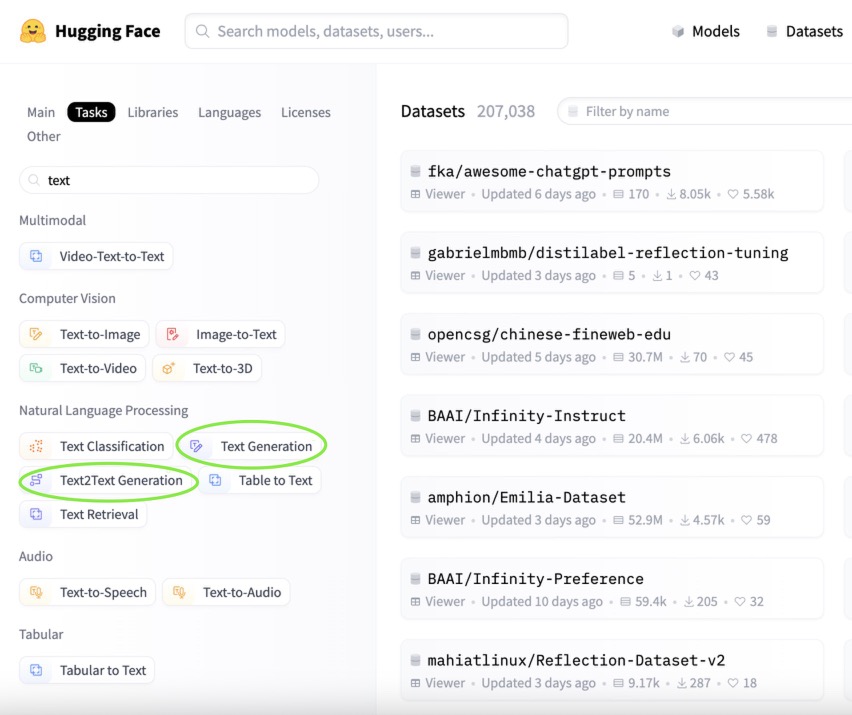}
    \caption{Task selection labels on the Huggingface portal showing NLP/ text category. The highlighted labels ``Text Generation'' and ``Text2Text Generation'' satisfied our needs among the listed tasks for code generation/ completion and their associated datasets.
    } 
    \label{fig:taskslection}
\end{figure*}

\textbf{Identifying adversarial attack generation tools:}
\edit{Based on these 338 GitHub projects, we applied a systematic filtering process to ensure relevance to our study. First, we examined the first 5 pages of search results and removed projects without an accompanying paper, as our study required well-documented methodologies for reproducibility. Next, we read the abstract and introduction of each accompanying paper to assess alignment with our research focus on perturbation techniques for robustness evaluation. We then reviewed project files, removing those lacking datasets or incompatible with our experimental environment (assessed through README.md and requirements.txt). Through this rigorous manual review, we refined our selection to 10 projects, which were further evaluated. \minor{We selected ReCode and OpenAttack from ten initially reviewed tools based on criteria including perturbation granularity, input compatibility (code + text), public accessibility, and extensibility. ReCode provides deterministic, code-specific perturbations, while OpenAttack supports diverse natural language attacks applicable to task descriptions and comments. The remaining tools were excluded due to limited applicability, lack of granularity, or poor documentation.}}

\textbf{Identifying Code Datasets:} 
\edit{Based on our exploratory analysis, we applied a structured dataset evaluation process to ensure robustness and comparability in our study. We first removed datasets unrelated to code completion tasks. Next, we prioritized datasets with widespread adoption in robustness research to maintain test reliability. Given our focus on Python-based models, we filtered out datasets containing other programming languages to ensure consistency. Additionally, we evaluated dataset complexity and benchmark availability, selecting datasets with existing evaluation results to facilitate direct performance comparison. To avoid redundancy, we excluded highly similar datasets such as MBJP and MBJSP, which closely resemble MBPP. As a result, we finalized two datasets, ensuring a diverse and reliable foundation for evaluating LLM4Code robustness under adversarial conditions.}

\textbf{Identifying Subject Models:} 
\edit{Based on our exploratory analysis (Step 1. Fig.~\ref{fig:timeline}), we systematically filtered models from 157,121 down to 5 model families through a multi-stage process:
\dcircled{1}\textbf{Code-Relevance Filtering:} We first removed non-Large Language models for Code, retaining 38,251 models.
\dcircled{2}\textbf{Original Developer Selection:} We excluded fine-tuned and derivative versions, keeping only models from their original creators, narrowing the set to 586 models.
\dcircled{3}\text{Size Constraints:} Due to experimental limitations, we focused on models smaller than 8B, reducing the selection to 6 model families: Codegen, Incoder, GPT-J, CodeT5, CodeLlama, and Llama3.
\dcircled{4}\textbf{Final Selection:} The CodeT5 model was excluded from our robustness evaluation due to consistently invalid test results, with all evaluation metrics yielding zero. Consequently, our final selection focused on five model families. This anomaly may also explain why the prior study~\cite{wang2022recode} ReCode similarly omitted CodeT5 from their reported findings, despite including it in their replication package.
Thus, the structured filtering ensured that our selected models were representative of modern LLM4Code, computationally feasible, and aligned with robustness evaluation benchmarks.}

\edit{Our filtering of both models and datasets was guided by methodological rigor, ensuring the final benchmark was reproducible, task-relevant, and aligned with prior robustness studies. Models were chosen based on applicability to code generation or completion, widespread use, support for Python, and prior inclusion in benchmarking efforts.}

\subsection{Experimental Setup} 
In our study's experimental setup~\cite{replication}, we employed two distinct servers to concurrently execute the models. The first server had an Nvidia V100 GPU and utilized Python version 3.8 and 3.10. In contrast, the second server operated with an Nvidia RTX A6000 and was configured using a conda-built virtual environment, also running Python 3.8. 
Python 3.8 is for replications and Python 3.10 is required by Llama models.

\subsection{Replication of ReCode}\label{sec:rep-recode} 
In this study, we replicated the ReCode project~\cite{wang2022recode} to validate the consistency of the reported results across different contexts, such as datasets, models, and environmental conditions. The replication aims to ensure the reliability of the original findings, ensuring that the results are not influenced merely by randomness, as recommended by Wohlin et al.~\cite{wohlin2012experimentation}. Thus, we meticulously followed the methodologies outlined in~\cite{wang2022recode} to validate and compare the robustness of various Large Language models for Code against adversarial attacks. We chose to replicate the ReCode project precisely because of its comprehensive approach to evaluating the vulnerability of programming language models to targeted adversarial attacks. This project provided a well-documented framework and benchmarks essential for establishing a reliable comparison and ensuring that our findings are consistent with established results in the field.

\subsection{Code Completion Dataset Construction}\label{sec:construction}
We leveraged existing benchmark datasets known for their relevance in code synthesis and execution tasks. As described in \ref{sec:overview}, we choose the HumanEval~\cite{chen2021evaluating} and MBPP (Mostly Basic Python Problems)~\cite{austin2021program} datasets, which are well-established in the domain of automated code generation evaluation.

\textbf{HumanEval Dataset}: HumanEval~\cite{chen2021evaluating} is a dataset consisting of a collection of 196 Python programming problems, 
each accompanied by a function signature, a docstring describing the task, test cases, and a correct reference implementation. This dataset is primarily used for evaluating Large Language models for Code' capabilities in synthesizing function bodies that pass associated unit tests. It was introduced by OpenAI as part of the Codex evaluation and has been widely adopted for benchmarking code completion and synthesis models.



\textbf{MBPP Dataset}: Similarly, we utilized the MBPP dataset~\cite{austin2021program}, which consists of a collection of Python programming problems designed to evaluate Large Language models for Code. This dataset includes 974 programming tasks, each with a natural language prompt describing the task and corresponding Python code solutions. MBPP covers a wide range of real-world coding challenges, from simple functions to more complex algorithms, ensuring a comprehensive assessment of the models' capabilities. It also provides test cases for each problem, allowing for automated evaluation of the generated code's correctness. MBPP is widely recognized as a valuable benchmark for testing the practical robustness of Large Language models for Code in generating accurate and functional Python code based on natural language descriptions.

\edit{In the datasets, as shown in listing~\ref{lst:human_eval}, each task contains the fields of ``task\_id'',  ``prompt'',``entry\_point'', ``canonical\_solution'', and ``test''. The prompt is task description of the input, entry\_point is the function name, and canonical\_solution is the expected outputs. The examples are shown in Section~\ref{sec:rq}. There are 62\% of all the perturbed datasets including only task descriptions, and 38\% including code and task descriptions.
Listing~\ref{lst:result_unperturbed} and listing~\ref{lst:result_perturbed} are the files of result to evaluation output. Listing~\ref{lst:result_unperturbed} shows the output for an unperturbed task input and listing~\ref{lst:result_perturbed} is the output for a perturbed input (FuncRenameInflectionalVariation) dataset. We define the correct output of a code generation model as follows:
\begin{itemize}
\item Task ID: A unique identifier for the coding problem (e.g., "HumanEval/0").
\item Completion: The actual generated output of the model.
\item Input: The initial function signature and docstring and test cases which is used to evaluate the correctness of the generated code.
\item Original Prediction: The expected, correct code output.
\item Result: Show the pass or the error information
\item Mean\_logp: In our results, all the Mean\_logp is None
\item Pass/Fail Status: Whether the generated code successfully passes test cases.
\end{itemize}
}

\noindent\begin{minipage}{.95\textwidth}
\begin{lstlisting}[language=Python, label=lst:human_eval, caption=An example showing the structure of output of humaneval dataset.]
{"task_id": "HumanEval/0",

"prompt": "from typing import List\n\n\n
def has_close_elements(numbers: List[float], threshold: float) -> bool:\n
\"\"\" Check if in given list of numbers, are any two numbers closer to each other than\n
given threshold.\n
>>> has_close_elements([1.0, 2.0, 3.0], 0.5)\n
False\n
>>> has_close_elements([1.0, 2.8, 3.0, 4.0, 5.0, 2.0], 0.3)\n
True\n    \"\"\"\n",

"entry_point": "has_close_elements",

"canonical_solution":
"    for idx, elem in enumerate(numbers):\n
       for idx2, elem2 in enumerate(numbers):\n
            if idx != idx2:\n
                distance = abs(elem - elem2)\n
                if distance < threshold:\n
                    return True\n\n
     return False\n", 
     
"test": "\n\nMETADATA = {\n    'author': 'jt',\n    
'dataset': 'test'\n}\n\n\n
def check(candidate):\n
    assert candidate([1.0, 2.0, 3.9, 4.0, 5.0, 2.2], 0.3) == True\n   
    assert candidate([1.0, 2.0, 3.9, 4.0, 5.0, 2.2], 0.05) == False\n 
    assert candidate([1.0, 2.0, 5.9, 4.0, 5.0], 0.95) == True\n    
    assert candidate([1.0, 2.0, 5.9, 4.0, 5.0], 0.8) == False\n    
    assert candidate([1.0, 2.0, 3.0, 4.0, 5.0, 2.0], 0.1) == True\n   
    assert candidate([1.1, 2.2, 3.1, 4.1, 5.1], 1.0) == True\n    
    assert candidate([1.1, 2.2, 3.1, 4.1, 5.1], 0.5) == False\n\n"}
\end{lstlisting}
\end{minipage}\hfill

\noindent\begin{minipage}{.45\textwidth}
\begin{lstlisting}[escapechar=!, language=Python, label=lst:result_unperturbed, caption=An example showing the structure of result of unperturbed dataset.]
"task_id": "HumanEval/0",
"completion": "
!\colorbox{yellow}{for i in range(len(numbers) - 1):}!\n
    !\colorbox{yellow}{if numbers[i] - numbers[i + 1] < threshold:}!\n
        !\colorbox{yellow}{return True}!\n
!\colorbox{yellow}{return False}!",
  
"ori_pred": "
for i in range(len(numbers) - 1):\n
    if numbers[i] - numbers[i + 1] < threshold:\n
        return True\n
return False\n\n\n",
  
def !\colorbox{yellow}{has}!_close_elements_2(numbers: List[float], threshold: float) -> bool:\n

......

"Result": "failed:",
"Passed:" false
\end{lstlisting}
\end{minipage}\hfill
\noindent\begin{minipage}{.45\textwidth}
\begin{lstlisting}[escapechar=!, language=Python, label=lst:result_perturbed, caption=An example showing the structure of result of one `perturbation(FuncRenameInflectionalVariation)' dataset.]
"task_id": "HumanEval/0",
"completion": "
!\colorbox{yellow}{for i in range(len(numbers) - 1):}!\n
    !\colorbox{yellow}{for j in range(i + 1, len(numbers)):}!\n
        !\colorbox{yellow}{if abs(numbers[i] - numbers[j]) < threshold:}!\n
            !\colorbox{yellow}{return True}!\n
!\colorbox{yellow}{return False}!",
  
"ori_pred": "
for i in range(len(numbers) - 1):\n
    if numbers[i] - numbers[i + 1] < threshold:\n
        return True\n
return False\n\n\n",
  
def !\colorbox{yellow}{had}!_close_elements_2(numbers: List[float], threshold: float) -> bool:\n

......

"Result": "failed:",
"Passed:" false
\end{lstlisting}
\end{minipage}

\subsection{LLM4Code (Large Language models for Code)} 

Pre-trained models have significantly advanced the fields of Natural Language Processing (NLP)~\cite{yuan2019adversarial} and code-level machine learning, demonstrating notable performance improvements across various tasks. As described in \ref{sec:overview}, we choose five pre-trained models/families, including the CodeGen~\cite{nijkamp2022codegen}, InCoder~\cite{fried2022incoder}, and GPT-J~\cite{de2022zero} families, CodeLlama, and Llama-3 
to evaluate Large Language models for Code' robustness under different types of adversarial attacks. The selection of these models was intentional, 
\edit{Our study focuses on evaluating LLM4Code primarily in the context of code completion tasks. While our perturbation strategies touch on elements such as code generation prompts, partial function definitions, and natural language descriptions, the core evaluation remains within the code completion domain. This targeted assessment provides insight into LLM4Code robustness under adversarial conditions, reflecting realistic scenarios encountered in software development workflows.}

\textbf{CodeGen}~\cite{nijkamp2022codegen} models, developed by Salesforce, are autoregressive language models trained on a mixture of natural and programming languages. Designed to assist in code generation, CodeGen models act as AI pair programmers, capable of generating syntactically and semantically correct code. 
They have been fine-tuned on diverse programming languages and excel in tasks such as code translation, completion, and documentation. 
The CodeGen family includes models of various sizes (including CodeGen-2B-mono, CodeGen-2B-multi, CodeGen-6B-mono, CodeGen-6B-multi, CodeGen-16B-mono, and CodeGen-16B-multi) with the CodeGen-16B achieving state-of-the-art performance on numerous benchmarks by building upon the success of earlier models such as GPT-3. 

\textbf{InCoder}~\cite{fried2022incoder} model family, including models such as PolyCoder, was developed by Facebook AI and other researchers. These transformer-based models are pre-trained on a large GitHub code corpus and are known for their bidirectional context understanding, allowing them to complete code by considering both preceding and following contexts. This feature makes InCoder models particularly effective for tasks such as code repair and bug fixing. 

\textbf{GPT-J}~\cite{de2022zero}, developed by EleutherAI, is a large-scale, autoregressive language model trained on a blend of natural and programming languages. GPT-J functions as a versatile AI coding assistant that is adept at understanding complex contexts to generate syntactically accurate and semantically rich code. Fine-tuned across a wide array of programming languages, GPT-J is capable of performing various coding tasks, including code translation, completion, and documentation. It represents an evolution from earlier models such as GPT-3, with specific enhancements to handle the intricacies of code syntax and semantics, making it a powerful tool for developers seeking to streamline the coding process and boost productivity.

\textbf{CodeLlama} is particularly adaptable to diverse programming environments, excelling in tasks such as code translation, completion, and detailed documentation. \textbf{Llama-3} extends these capabilities further by incorporating meta-learning techniques, enabling rapid adaptation to new programming languages and tasks with minimal additional training. Both models are trained on extensive datasets encompassing a wide range of programming and natural languages, allowing them to perform effectively across various coding scenarios. 

\subsection{Classification of Attacks for Large Language models for Code} 

\edit{In this study, we systematically evaluate the robustness of Large Language Models for Code (LLM4Code) under various adversarial perturbations, and we aim to provide insights that may inform future efforts to enhance model robustness.} We get idea from OpenAttack~\cite{zeng2020openattack} which provided the classification of perturbations in different levels. We introduce this classification method to code-specific perturbations because the similarity of code and natural language which we discussed before. We scrutinize the pre-perturb dataset and perturbed datasets to understand the characteristics of these perturbations on these samples. Based on it, we propose a novel two-dimensional classification system. This classification categorizes adversarial attacks based on both content type and granularity, providing a comprehensive framework to assess vulnerabilities in LLM4Code. The first dimension in our classification framework distinguishes three key content types that are targeted by adversarial attacks: 
\dcircled{1} \textbf{Code:} \textit{Direct perturbations in the input code itself, such as changing the structure or logic of the code}. \dcircled{2} \textbf{Task Descriptions:} \textit{Adversarial modifications in the natural language descriptions provided to the model as prompts guide code generation}. \dcircled{3} \textbf{Comments:} \textit{Perturbations introduced into code comments, which, while not directly affecting code execution, can mislead the developer or hinder model understanding}.

The second dimension categorizes the attacks by their level of granularity: \dcircled{1} \textbf{Character-level perturbations} target individual characters within the code, task description, or comments. These subtle changes, such as typographical errors or minor character substitutions, can have far-reaching consequences, leading to cascading errors in code compilation or execution. \dcircled{2}\textbf{ Word-level perturbations} involve altering or replacing entire words or tokens, which can disrupt the semantic meaning of both code and natural language components, affecting the model’s ability to generate correct or coherent outputs. \dcircled{3}\textbf{ Sentence/statement-level perturbations} introduce changes at a macroscopic level, modifying entire logic blocks or sentences in task descriptions, comments, or code. These changes can lead to significant structural or logical errors, potentially causing the model to generate faulty or vulnerable code.

We evaluated these approaches to understand how different attacks affect Large Language models for Code at various content types and levels of granularity. For example, OpenAttack offers perturbation strategies focused on the task description, while ReCode provides pre-perturbed datasets with more generalized attacks on code and comments. The classification system is designed to comprehensively categorize the types of adversarial attacks that Large Language models for Code may encounter. The granularity and content types of specific perturbations used in our study are detailed in Tables~\ref{tab:reclassification},~\ref{tab:reclassification_for_RQ2},~\ref{tab:reclassification_for_RQ2.2},~\ref{tab:reclassification_for_RQ3} in section~\ref{sec:rq}, demonstrating the varied effects these attacks can have on model robustness.

This two-dimensional classification allows researchers and practitioners to assess the vulnerabilities of LLM4Code systematically. Understanding the impacts of specific perturbations across different content types and levels of granularity will guide the development of more resilient and reliable code-generation tools in real-world applications. This structured evaluation serves as a foundation for improving the security and reliability of AI-driven coding solutions in modern software development.

\subsection{Application of Perturbations}
For both datasets (HumanEval and MBPP), we introduced adversarial perturbations at the character, word, and sentence levels using tools such as ReCode~\cite{wang2022recode} and OpenAttack~\cite{zeng2020openattack}. We use ReCode~\cite{wang2022recode} to generate perturbed datasets for both task descriptions and code portions, ensuring meaningful perturbations across different content types. 
OpenAttack provided mechanisms to perturb the task description parts of the datasets, focusing on natural language descriptions to simulate real-world conditions where inputs may contain adversarial attacks. Openattack also classify these attack models into different levels, such as character, word, and sentence level. Because of it, we use OpenAttack to enrich the number of our perturbed datasets for task descriptions. \minor{In ReCode, perturbation targets are generated using fixed rule-based templates with deterministic execution. OpenAttack employs stochastic techniques, but outcomes are reproducible when executed with a fixed random seed and consistent configuration.}
\edit{For our perturbations or attacks, we utilized established adversarial robustness frameworks: ReCode and OpenAttack, along with a custom perturbation for comment insertion (randominsertcomments). Every sample in the dataset was perturbed and included in the evaluation.}
\edit{\begin{itemize}
    \item \textbf{Non-code} perturbations targeted \textit{natural language components} (e.g., task descriptions, comments) and included \textbf{word substitutions, grammar modifications, and syntactic alterations}~\cite{zeng2020openattack}. These perturbations simulate realistic \textbf{adversarial inputs} that a developer or hacker might introduce through natural language modifications.
    \item \textbf{Code} perturbations modified \textbf{code structures}, affecting \textbf{function renaming, variable renaming, and syntax errors}~\cite{wang2022recode}. Unlike OpenAttack, ReCode does not classify perturbations by task descriptions or granularity levels.
    \item \textbf{Attack Application Methodology:}
    \begin{itemize}
        \item For \textbf{Code perturbations}, each input was perturbed \textbf{once per sample, affecting a single target type} (e.g., all variable names in a sample were consistently modified).
        \item For \textbf{Non-code perturbations}, each input was perturbed \textbf{once per sample, affecting multiple elements} (e.g., function names and task descriptions simultaneously).
        \item Some perturbations, particularly those on \textbf{task descriptions}, used \textit{random selection methods} (NLTK-based) to introduce \textbf{linguistic variation}. In contrast, function and variable renaming perturbations were \textbf{deterministic}, ensuring consistency.
    \end{itemize}
\end{itemize}}
\edit{The definitions of all perturbations can be found in table~\ref{tab:reclassification}, table~\ref{tab:reclassification_for_RQ2}, table~\ref{tab:reclassification_for_RQ2.2} and table~\ref{tab:reclassification_for_RQ3} (Section \ref{sec:rq}), where we categorize perturbations at different levels of granularity and provide representative examples. Additionally, Section \ref{sec:rq} includes detailed examples illustrating how each perturbation type modifies code completion tasks.}

\edit{The perturbations applied in this study, including synonym replacements, follow the dataset released in prior work~\cite{wang2022recode}~\cite{zeng2020openattack}. While most word substitutions maintain semantic consistency, some cases—such as 'moment' vs. 'here and now'—may introduce slight variations in meaning. Our robustness analysis accounts for these variations by evaluating multiple perturbation types rather than relying solely on individual word replacements.}
In addition, these two datasets, used extensively for code generation model evaluations, were originally developed and reviewed by their respective research teams at OpenAI and Google. These datasets have been widely adopted by the research community for benchmarking purposes. 

\edit{Our perturbations were designed to maintain original semantics in task descriptions and preserve functional correctness in code modifications. While we conducted spot-checking of samples, a structured manual validation process involving multiple authors could further strengthen result reliability.}

As a result, the datasets are considered robust and reliable for testing. Their widespread usage and peer-review in the academic community have established them as standard resources for evaluating model performance and robustness under various conditions. By incorporating these adversarial perturbations, we constructed a comprehensive dataset that tests not only the models' abilities to generate syntactically and semantically correct code but also their resilience to adversarial inputs.

\edit{
\subsection{Manual Validation of Perturbation Semantics}
To assess the semantic integrity of perturbed samples, we conducted a manual validation on a statistically representative subset of the full perturbation dataset. This dataset, derived from the HumanEval and MBPP benchmarks, comprises a total of 38,692 perturbed instances. These were generated using 28 ReCode perturbation methods and 6 OpenAttack methods across the original datasets: 164 HumanEval and 974 MBPP samples. Table~\ref{tab:perturbation-counts} provides a breakdown of this population.
\begin{table}[h]
\centering
\caption{Distribution of Perturbed Dataset Instances}
\label{tab:perturbation-counts}
\begin{tabular}{|l|c|c|}
\hline
\textbf{Dataset} & \textbf{ReCode (28 perturbations)} & \textbf{OpenAttack (6 perturbations)} \\
\hline
HumanEval & 4592 (164 \texttimes 28) & 984 (164 \texttimes 6) \\
\hline
MBPP & 27272 (974 \texttimes 28) & 5844 (974 \texttimes 6) \\
\hline
\textbf{Total Instances} & \multicolumn{2}{c|}{\textbf{38,692}} \\
\hline
\end{tabular}
\end{table}
Using a 95\% confidence level and a 5\% margin of error, we calculated a required sample size of 381 for validation. Samples were drawn using stratified random sampling to ensure proportional representation across both datasets and perturbation tools. The proportions and resulting sample counts are summarized in Table~\ref{tab:sample-counts}.
\begin{table}[h]
\centering
\caption{Sample Counts by Dataset and Perturbation Tool}
\label{tab:sample-counts}
\begin{tabular}{|l|c|c|}
\hline
\textbf{Dataset} & \textbf{ReCode} & \textbf{OpenAttack} \\
\hline
HumanEval & 43 & 12 \\
\hline
MBPP & 255 & 70 \\
\hline
\textbf{Total} & \textbf{298} & \textbf{82} \\
\hline
\end{tabular}
\end{table}
To maintain consistency across different perturbation types, we distributed the samples evenly among the perturbation levels within each tool. For ReCode, the 28 methods include 8 character-level (28.6\%), 9 word-level (32.1\%), and 11 sentence-level (39.3\%) perturbations. For OpenAttack’s 6 methods, the breakdown includes 1 character-level (16.7\%), 3 word-level (50.0\%), and 2 hybrid character/word-level (33.3\%) perturbations.
All 381 samples were manually reviewed by the authors to verify whether the perturbations preserved the semantic intent of the original code or task description. The validation results revealed that ReCode perturbations generally maintained semantic fidelity and readability. In contrast, OpenAttack perturbations, particularly those involving synonym replacement and character-level manipulations, were found to be more aggressive and occasionally degraded the semantic coherence of the input. A notable example is the DeepWordBug attack, which introduced Unicode substitutions (e.g., replacing the character \texttt{e} with \texttt{\textbackslash u0435}), leading to potential interpretation or display issues.\\
Overall, this manual validation confirmed that while most perturbations preserved semantics, aggressive adversarial strategies from OpenAttack introduced risks of semantic drift, underscoring the need for careful perturbation design in robustness evaluation.
}

\section{Evaluation} \label{sec:eval}


\edit{To evaluate model robustness systematically, we introduce a classification system that organizes datasets based on perturbation level(character, word, and sentence/statement levels), and perturbation targets(code, task description, and comments).. This classification differs from the original dataset categorization, which focused on tool source, and task type, ensuring a more structured assessment of model performance under adversarial conditions.
}

\subsection{Evaluation metrics}\label{sec:eval-metrics} 

\edit{
In this work, we use the robustness evaluation metrics introduced in prior research~\cite{chen2021evaluating,wang2022recode} to measure LLM4Code performance under adversarial perturbations. To ensure reproducibility and consistency, we control random seed selection as follows:
\begin{itemize}
 \item We initially tested multiple seed values (1, 3, 5, 7) and observed no significant differences in performance.
\item To maintain comparability with prior studies, we set seed = 5, aligning with ReCode's experimental design.
\item Experiments were repeated three times for two models, yielding identical results across all trials, confirming experimental stability.
\item Based on these findings, we did not repeat experiments for the remaining models, as previous trials demonstrated consistency.
\end{itemize}
This controlled approach ensures that our findings are stable, minimizing random variability while maintaining direct comparability with existing benchmarks.
}\\
\edit{Let k be the number of top outcomes, where a top outcome refers to one of the k best-ranked outputs generated by the model when prompted with a query. These rankings are typically based on model confidence, likelihood scores, or correctness evaluations. In code completion, top-k evaluation helps assess whether at least one of the k generated outputs meets the expected solution criteria.}
In our work, we set \(\textbf{k} = 1\), making the evaluation only regard to the top outcome. 
Then we evaluate the metrics presented below.

\noindent  
\textbf{Pass@k}~\cite{chen2021evaluating}, whose value range is [0, 1], is the probability for which the model generates the correct output within the top k outcomes. 

\noindent \textbf{RPs@k (Robust Pass@k)}~\cite{wang2022recode} 
measures the fraction of prompts in the dataset for which the model generates the correct output within the top k outcomes under adversarial perturbations.
A value of 0 means that the model fails to generate the correct output for any prompt under perturbations, while a value of 1 indicates that the model always generates the correct output. It is calculated according to Eq.~(\ref{eq:eq1}) presented below:

\[
RP_S@k= \mathbb{E}_x \left[ 1 - \frac{\binom{n - rc_s(x)}{k}}{\binom{n}{k}} \right] \tag{1}
\label{eq:eq1}\]\\
n is the total number of possible completions. $rc_s(x)$=$\sum_{i=1}^{n} c_{i,s}(x)$. $c_{i,s}$ means worst-case correctness across generated samples.
$c_{i,s}$=1 means generated samples are all correct, and $c_{i,s}$=0 means generated samples are all wrong.  
\edit{For clarity and consistency, we standardize the notation for our evaluation metrics throughout the paper. Specifically, Pass@k refers to the conventional correctness metric for LLMs for Code, measuring the probability that at least one of the top-k generated outputs passes all associated unit tests for a given (unperturbed) task. In contrast, RPs@k denotes the robustness-aware variant, which measures the probability that at least one of the top-k outputs passes all test cases derived from perturbed inputs. All occurrences of these terms in the main text, figures, and tables have been revised to reflect this standardized terminology.}

\noindent \textbf{RDs@k (Robust Drops@k)}~\cite{wang2022recode}, whose value range is [0, 1], quantifies the relative drop in Pass@k 
performance between worst-case scenarios (as measured by RPs@k) and no perturbation scenarios. 
A value of 0 means that there is no drop in performance between worst-case and average-case scenarios, while a value of 1 indicates a complete drop in performance, with no overlap between the two scenarios. It is calculated according to Eq.~(\ref{eq:eq2}) presented below:

\[
\text{RD}_s@k= \frac{\text{Pass@k} - \text{Robust Pass}_s@k}{\text{Pass@k}} \tag{2}
\label{eq:eq2}\]

\noindent \textbf{RRs@k (Robust Relatives@k)}~\cite{wang2022recode}, whose value range is [0, 1], measures the fraction of changes in model predictions between original prompts and perturbed prompts relative to the total size of the dataset. A value of 0 means that there are no changes in model predictions between original and perturbed prompts, while a value of 1 indicates that all model predictions change between the two scenarios. RRs@1 is calculated according to Eq.~(\ref{eq:eq3}) and RRs@k is calculated according to Eq.~(\ref{eq:eq4}) presented below:

\[
\text{RRs@1}= \frac{\text{RC}^{[+]}_s + \text{RC}^{[-]}_s}{N} \tag{3}
\label{eq:eq3}\]

\[
\text{RRs@k}= \mathbb{E}_x \left[ 2 - \frac{\binom{n - rc_s^{[-]}(x)}{k}}{\binom{n}{k}} - \frac{\binom{n - rc_s^{[+]}(x)}{k}}{\binom{n}{k}} \right] \tag{4}
\label{eq:eq4}\]\\
$RC_s^{[+]}$ represents the number of times the model's output changes positively (i.e., an improvement) between the original and the perturbed input for a seed (s). 
$RC_s^{[-]}$ represents the number of times the model's output changes negatively (i.e., a deterioration) between the original and perturbed inputs for a seed s.

In our assessment, we assign the highest importance to the Robust Pass metric (RPs@k), as it directly reflects the model's ability to successfully handle perturbations without failure. 
When models exhibit similar RPs@k, we then evaluate the Robust Drops metric (RDs@k), which reflects the extent of performance degradation. A lower Robust Drops value is preferable, as it signifies fewer instances of performance degradation. If models also show comparable results in terms of Robust Drops, we further differentiate their performance by examining the Robust Relatives metric (RRs@k). This metric provides a comparative analysis of the consistency of performance in different perturbations. By following this hierarchical evaluation approach, we ensure a comprehensive and nuanced assessment of the robustness of the model.

\subsection{Benchmark} 

\noindent\textbf{Benchmark for datasets.}
In this study, we experimented with the HumanEval~\cite{chen2021evaluating} and MBPP~\cite{austin2021program} datasets in our benchmarks. These datasets are particularly suited for our analysis due to their extensive use in previous research, which provides a plethora of comparative data. HumanEval, known for its challenging coding problems and test cases, allows an in-depth evaluation of a model’s code synthesis capabilities. 
\edit{Functional correctness, typically assessed through test case execution, is a standard in the field for evaluating Large Language models for Code (LLM4Code)~\cite{chen2021evaluating}~\cite{cao2024can}~\cite{hendrycks2021measuring}. Benchmarks such as HumanEval and APPS have been widely used to measure functional correctness and assess model performance in generating syntactically and semantically correct code.}
MBPP, with its focus on Python programming problems, serves as an excellent counterpart to HumanEval by providing a more specialized context for evaluating code generation within a single language's ecosystem. 
By employing these benchmark datasets, we ensure that our study's findings are grounded in the same context as prior work, allowing for a direct and fair comparison of model performance. This choice of datasets also aligns with the broader research standards in the fields of machine learning and software engineering 
for rigorous and standardized evaluation in code generation research.
\edit{As we mentioned in section\ref{sec:construction}, each dataset sample consists of structured fields, including task\_id, prompt, entry\_point, canonical\_solution, and test(shown in Table~\ref{tab:structured fields}. The prompt field provides the function signature and description, guiding model-generated code. The canonical\_solution contains the expected implementation, meanwhile “test” includes unit tests for correctness verification. This structure ensures a systematic evaluation of Large Language models for Code, where correctness is determined based on unit test execution. Listing~\ref{lst:promp} shows an example of prompt, and listing~\ref{lst:solution} shows an example of expected output.}

\begin{table*}[!ht]
\centering
\caption{The definitions of structured fields in datasets}
\label{tab:structured fields}
\resizebox{\textwidth}{!}{%
\begin{threeparttable}
    \begin{tabular}{l|l} 
\hline
\hline
\textbf{Field Name}         & \textbf{Definition}\\
\hline
\hline
task\_id & A unique identifier for each coding task (e.g., "HumanEval/0").\\
\hline
prompt & The input provided to the model, typically containing a function signature and a docstring describing the expected behavior.\\
\hline
entry\_point &The main function that needs to be implemented (e.g., "has\_close\_elements").\\
\hline
canonical\_solution & The correct or expected implementation of the function.\\
\hline
test & A set of unit tests used to evaluate whether the generated code produces the correct outputs.\\
\hline
\end{tabular}%
\end{threeparttable}%
}
\end{table*}

\noindent\begin{minipage}{.45\textwidth}
\begin{lstlisting}[language=Python, label=lst:promp, caption=An Example of Input (Prompt): The model receives a function signature and a description of its intended behavior.]
from typing import List
def has_close_elements(numbers: List[float], threshold: float) -> bool:
""" 
Check if in given list of numbers, are any two numbers closer to each other than given threshold.
>>> has_close_elements([1.0, 2.0, 3.0], 0.5)
False
>>> has_close_elements([1.0, 2.8, 3.0, 4.0, 5.0, 2.0], 0.3)
True    
"""
\end{lstlisting}
\end{minipage}\hfill
\noindent\begin{minipage}{.45\textwidth}
\begin{lstlisting}[language=Python, label=lst:solution, caption=An Example of Expected Output (Canonical Solution): The correct implementation of the function.]
    for idx, elem in enumerate(numbers):
       for idx2, elem2 in enumerate(numbers):
            if idx != idx2:
                distance = abs(elem - elem2)
                if distance < threshold:
                    return True
     return False, 
\end{lstlisting}
\end{minipage}\hfill


\noindent\textbf{Subject models.}
\edit{We evaluate five model families (CodeGen, InCoder, GPT, CodeLlama2, and Llama3). Within the CodeGen family, we include two variants—CodeGen-2B-Mono and CodeGen-2B-Multi— bringing the total number of models evaluated to six. This distinction ensures a comprehensive assessment of robustness across different architectures and configurations.}These models have been previously evaluated under various perturbation methods, establishing a benchmark for our approach. Each baseline has been tested against adversarial conditions, with performance metrics such as code generation or completion accuracy, error rates, and robustness to syntactic and semantic anomalies reported in peer-reviewed studies~\cite{wang2022recode}. Our comparative analysis re-evaluates these baselines using the same high-success-rate perturbation methods from ReCode and OpenAttack that we applied in our study, ensuring consistent and fair comparisons. Additionally, we introduce our novel adversarial attack classification system, systematically applying character-level, word-level, and sentence-level perturbations to assess the impact on each model. We focus on how performance metrics degrade with increasing perturbation sophistication.

\edit{\noindent\textbf{Semantic Impact Levels.} In our analysis, we categorize perturbations based on their semantic impact levels rather than intensity, which may have been an unclear term. We define three levels of impact:
\begin{itemize}
\item Low Impact (Minimal Change): Alterations such as case changes in function/variable names\footnote{\label{fn3}The function/variable names are changed consistently (i.e., changing all occurrences of the same function/variable names.)}, which should not significantly affect model outputs.
\item Medium Impact (Semantic Shift): Replacing function/variable\footref{fn3} names with meaningful synonyms, which may challenge LLM4Code generalization.
\item High Impact (Disruptive Change): Replacing function/variable\footref{fn3} names with arbitrary or nonsensical words, which can severely degrade LLM4Code performance.
\end{itemize}
Our findings indicate that even low-impact perturbations can cause performance degradation, suggesting that LLM4Code are more sensitive to syntactic cues than expected. The full categorization of perturbations and their definitions can be found in Tables \ref{tab:reclassification},~\ref{tab:reclassification_for_RQ2},~\ref{tab:reclassification_for_RQ2.2}, and~\ref{tab:reclassification_for_RQ3}(Section \ref{sec:rq}).}


\subsection{Research questions (RQs)}\label{sec:rq}
\noindent\textbf{\rqone}

\noindent\minor{\textbf{Scope:} Perturbations are injected into the function signature and implementation of the code snippet that constitutes the prompt.}

\noindent\textbf{Objective:}
This research question explores the effects of introducing perturbations at various levels—character, word, and statement levels—directly within the code on the functional correctness of the output generated by Large Language models for Code. The aim is to quantify and understand how alterations are made to the code (including but not limited to function names and variable names) in datasets, instead of the task description (\textit{discussed in RQ2}), influencing the model's ability to generate syntactically correct and semantically accurate code. Through this investigation, we intend to shed light on the robustness of Large Language models for Code against a spectrum of adversarial modifications embedded within the code, assessing their capacity to maintain or restore the intended functionality of the code amidst such perturbations. The results of this analysis will provide valuable insights into the vulnerabilities of current code generation technologies to code-level perturbations and guide the enhancement of models for greater security and reliability in automated coding processes.

\noindent\textbf{Approach:}
We use the perturbation methods outlined in ReCode to create perturbed datasets for HumanEval and MBPP. These datasets are then reclassified based on the granularity of the perturbations: character-level, word-level, and statement-level. \edit{When perturbing datasets, we preserve natural semantics in task descriptions to ensure meaningful test conditions. However, for function and variable name perturbations, we do not enforce semantic constraints. We argue that LLMs should not rely on identifier names to generate functionally correct code, as this could introduce serious robustness and security risks. For instance, if an attacker modifies function names, leading to incorrect or unreliable code completion, it exposes a fundamental weakness in LLM4Code. A similar approach has been used in prior works~\cite{wang2022recode}~\cite{yang2022natural}, where variable names were perturbed using both natural and unnatural replacements. Extending this, we assess LLM4Code under function name perturbations, further exploring their robustness beyond variable renaming.}The classification and definition of each perturbation are detailed in Table~\ref{tab:reclassification}. \edit{This classification is done to evaluate different models comprehensively using these distinct datasets.} The final phase involves a thorough assessment of models using these datasets, which helps in extracting three primary metrics: RPs@k (Robust Pass@k), RDs@k (Robust Drops@k), and RRs@k (Robust Relatives@k), as explained in 
Section~\ref{sec:eval-metrics}. We provide some examples below ---Listing~\ref{lst:original} is an example without perturbations, while Listings~\ref{lst:changechar}~\ref{lst:RN}~\ref{lst:loop} shows examples with different perturbations. Listing~\ref{lst:changechar} demonstrates a change in the function name from `rolling\_max' to `rol\textbf{L}ing\_\textbf{M}ax.' Listing~\ref{lst:RN} shows the transformation of the variable `running max' into `wrNLwMt6s5h', and Listing~\ref{lst:loop} presents the transformation of a `\textbf{for}' loop into a `\textbf{while}' loop. \edit{To ensure statistical validity, we applied the Wilcoxon signed-rank test to compare model performance under different perturbation conditions and used Bonferroni correction to adjust for multiple comparisons.}

\noindent\begin{minipage}{.45\textwidth}
\begin{lstlisting}[language=Python, label=lst:original, caption=Unperturbated code. Subsequent Listings~\ref{lst:changechar}~\ref{lst:RN}~\ref{lst:loop} are perturbated at different granularity levels from this code derivative.]
def rolling_max(numbers):
    """
    From a given list of integers, generate a list of rolling maximum element found until given moment in the sequence.
    >>> rolling_max([1, 2, 3, 2, 3, 4, 2])
    [1, 2, 3, 3, 3, 4, 4]
    """
    running_max = None
    result = []

    for n in numbers:
        if running_max is None:
            running_max = n
        else:
            running_max = max(running_max, n)
        result.append(running_max)
    return result
\end{lstlisting}
\end{minipage}\hfill
\begin{minipage}{.5\textwidth}
\begin{lstlisting}[escapechar=!,language=Python, label=lst:changechar, caption=Character-level perturbations; FuncRenameChangeChar. Two letters in the function name are changed from `rolling\_max' to `rolLing\_Max']
def !\colorbox{yellow}{rolLing\_Max}!(numbers):
    """
    From a given list of integers, generate a list of rolling maximum element found until given moment in the sequence.
    >>> !\colorbox{yellow}{rolLing\_Max}!([1, 2, 3, 2, 3, 4, 2])
    [1, 2, 3, 3, 3, 4, 4]
    """
    running_max = None
    result = []

    for n in numbers:
        if running_max is None:
            running_max = n
        else:
            running_max = max(running_max, n)
        result.append(running_max)
    return result
\end{lstlisting}
\end{minipage}

\noindent\begin{minipage}{.45\textwidth}
\begin{lstlisting}[escapechar=!,language=Python, label=lst:RN, caption=An example of word-level perturbations known as VarRenamerRN showing the variable `running\_max' that is altered to `wrNLwMt6s5h']
def rolling_max(numbers):
    """
    From a given list of integers, generate a list of rolling maximum element found until given moment in the sequence.
    >>> rolling_max([1, 2, 3, 2, 3, 4, 2])
    [1, 2, 3, 3, 3, 4, 4]
    """
    wrNLwMt6s5h = None
    result = []
    for n in numbers:
        if !\colorbox{yellow}{wrNLwMt6s5h}! is None:
            !\colorbox{yellow}{wrNLwMt6s5h}! = n
        else:
            !\colorbox{yellow}{wrNLwMt6s5h = max(wrNLwMt6s5h, n)}!
        result.append(wrNLwMt6s5h)
    return result
\end{lstlisting}
\end{minipage}\hfill
\begin{minipage}{.5\textwidth}
\begin{lstlisting}[escapechar=!, language=Python, label=lst:loop, caption=Example of statement-level perturbations named ForWhileTransformer. The `for' loop is changed to a `while' loop.]
def rolling_max(numbers):
    """
    From a given list of integers, generate a list of rolling maximum element found until given moment in the sequence.
    >>> rolling_max([1, 2, 3, 2, 3, 4, 2])
    [1, 2, 3, 3, 3, 4, 4]
    """
    running_max = None
    result = []
    index = 0
    !\colorbox{yellow}{while index < len(numbers):}!
        n = numbers[index]
        if running_max is None:
            running_max = n
        else:
            running_max = max(running_max, n)
        result.append(running_max)
        index += 1
    return result
\end{lstlisting}
\end{minipage}

\begin{table*}[!ht]
\centering
\caption{Classifying code perturbations into three classes: Character-level(C), Word-level(W), and Statement-Level(S)}
\label{tab:reclassification}
\resizebox{\textwidth}{!}{%
\begin{threeparttable}
    \begin{tabular}{l|l|l} 
\hline
\hline
\textbf{Perturbation Method }         & \textbf{Definition }  &\textbf{ Classification}  \\
\hline
\hline
FuncRenameSwapChar \textbf{(C1) }         & Swapping characters in function names.                                 & Character-Level\\
\hline
FuncRenameChangeChar \textbf{(C2)}        & Changing individual characters in function names.                     & Character-Level\\ 
\hline
tab\_indent \textbf{(C3) }                 & Modifying tab or space characters used for indentation.                           & Character-Level\\ 
\hline
split\_lines \textbf{(C4) }                & Altering line breaks which could involve changing character-level line break symbols. & Character-Level\\ 
\hline
FuncRenameCamelCase \textbf{(W1)}        & Modifying function names which could involve changing entire words if the function names are compound words.         & Word-Level \\
\hline
FuncRenameButterFinger \textbf{(W2)}      & This could involve typos in function names, affecting entire words.               & Word-Level \\ 
\hline
FuncRenameSynonymSub \textbf{(W3)}       & Substituting parts of function names with synonyms, changing whole words.         & Word-Level \\
\hline
VarRenamerCB \textbf{(W4)}                & Changing variable names, likely affecting entire words.                           & Word-Level \\ 
\hline
VarRenamerNaive \textbf{(W5) }            & Similar to VarRenamerCB, changing entire variable names.                          & Word-Level \\ 
\hline
VarRenamerRN \textbf{(W6)}                & Renaming variables, affecting whole words.                                        & Word-Level \\ 
\hline
FuncRenameInflectionalVariation \textbf{(S1)} & Changing function names by altering their inflection, which could affect the meaning of entire sentences or code statements.      &Statement-Level \\
\hline
DeadCodeInserter \textbf{(S2)}           & Inserting non-functional code, affecting sentence or statement structure.         & Statement-Level\\ 
\hline
ForWhileTransformer \textbf{(S3)}        & Transforming for loops to while loops, or vice versa, affecting entire code constructs. & Statement-Level\\
\hline
OperandSwap \textbf{(S4)}                 & Swapping operands in expressions, impacting the logic of statements.              & Statement-Level\\ 
\hline
new\_lines\textbf{ (S5) }                  & Inserting new lines, which can change the code structure at the sentence/statement level. & Statement-Level\\ 
\hline
new\_line\_aftercode \textbf{(S6)}          & Adding new lines after code statements, affecting the overall structure of the code. & Statement-Level\\ 
\hline
\end{tabular}%
\end{threeparttable}%
}
\end{table*}

\begin{table*}[!ht]
    \caption{Evaluation of Robustness Metrics on Three Code-Task Models Using the Humaneval Dataset Perturbed in the Code. An upward arrow ($\uparrow$) indicates that a higher value corresponds to better robustness, while a downward arrow ($\downarrow$) indicates the opposite. Meta-Llama-3-8B-Instruct shows the best performance} 
    \label{tab:results_RQ1_1}
    \resizebox{\textwidth}{!}{
    \begin{tabular}{ll|lll|lll|lll}
     \hline
       \textbf{Perturbations} & ~ & \multicolumn{3}{c}{\textbf{Meta-Llama-3-8B-Instruct}} & \multicolumn{3}{|c|}{\textbf{gpt-j-6b}} &\multicolumn{3}{c}{\textbf{CodeLlama-7b-hf}} \\ \hline
        \textbf{Level} & \textbf{Method} & \textbf{passatk}~\textcolor{blue}{$\uparrow$} &\textbf{drop(\%)}~\textcolor{red}{$\downarrow$} & \textbf{rel.(\%)}~\textcolor{red}{$\downarrow$} &\textbf{passatk}~\textcolor{blue}{$\uparrow$} & \textbf{drop(\%)}~\textcolor{red}{$\downarrow$} & \textbf{rel.(\%)}~\textcolor{red}{$\downarrow$} &\textbf{passatk}~\textcolor{blue}{$\uparrow$}& \textbf{drop(\%)}~\textcolor{red}{$\downarrow$} & \textbf{rel.(\%)}~\textcolor{red}{$\downarrow$}  \\ \hline
        
        \multirow{4}{*}{Character (\textbf{C})}
        ~ & C1 & 0.579 & 0.00 & 2.44 & 0.122 & -5.26 & 0.61 & 0.037 & 33.33 & 3.05 \\ \cline{2-11}
        ~ & C2 & 0.579 & 0.00 & 7.32 & 0.092 & 21.05 & 6.10 & 0.0310 & 44.44 & 2.44 \\ \cline{2-11}
        ~ & C3 & 0.652 & 1.83 & 7.32 & 0.244 & 2.44 & 7.93 & 0.561 & -196.77 & 42.07 \\ \cline{2-11}
        ~ & C4 & 0.652 & 1.83 & 9.76 & 0.244 & 2.44 & 4.27 & 0.189 & 0.00 & 1.22 \\ \cline{2-11}
        ~ & Median & 0.616 & 0.92 & 7.32 & 0.183 & 2.44 & 5.18 & 0.113 & 16.67 & 2.74 \\ \hline
    
        \multirow{6}{*}{Word (\textbf{W})} 
        ~ & W1 & 0.555 & 4.21 & 2.44 & 0.116 & 0.00 & 1.22 & 0.031 & 44.44 & 2.44 \\ \cline{2-11}
        ~ & W2 & 0.592 & -2.11 & 8.54 & 0.092 & 21.05 & 3.66 & 0.049 & 11.11 & 1.83 \\ \cline{2-11}
        ~ & W3 & 0.579 & 0.00 & 0.00 & 0.116 & 0.00 & 0.00 & 0.055 & 0.00 & 0.00 \\ \cline{2-11}
        ~ & W4 & 0.659 & 0.92 & 7.93 & 0.195 & 21.95 & 9.15 & 0.165 & 12.90 & 2.44 \\ \cline{2-11}
        ~ & W5 & 0.640 & 3.67 & 10.98 & 0.195 & 21.95 & 10.37 & 0.165 & 12.90 & 3.66 \\ \cline{2-11}
        ~ & W6 & 0.646 & 2.75 & 7.93 & 0.177 & 29.27 & 9.76 & 0.171 & 9.68 & 4.27$\backslash$ \\ \cline{2-11}
        ~ & Median & 0.616 & 1.83 & 7.93 & 0.146 & 21.50 & 6.40 & 0.110 & 12.01 & 2.44 \\ \hline
    
        \multirow{6}{*}{Statement (\textbf{S})} 
        ~& S1 & 0.585 & -1.05 & 3.05 & 0.110 & 5.26 & 1.83 & 0.042 & 22.22 & 1.22 \\ \cline{2-11}
        ~ & S2 & 0.518 & 22.02 & 23.17 & 0.140 & 43.90 & 15.85 & 0.085 & 54.84 & 10.37 \\ \cline{2-11}
        ~ & S3 & 0.646 & 2.75 & 6.71 & 0.213 & 14.63 & 9.76 & 0.122 & 35.48 & 7.93 \\ \cline{2-11}
        ~ & S4 & 0.671 & -0.92 & 7.93 & 0.207 & 17.07 & 9.15 & 0.183 & 3.23 & 1.83 \\ \cline{2-11}
        ~ & S5 & 0.659 & 0.92 & 6.71 & 0.226 & 9.76 & 7.32 & 0.213 & -12.90 & 6.10 \\ \cline{2-11}
        ~ & S6 & 0.659 & 0.92 & 7.93 & 0.183 & 26.83 & 9.15 & 0.189 & 0.00 & 2.44 \\ \cline{2-11}
        ~ & Median & 0.652 & 0.92 & 7.32 & 0.195 & 15.85 & 9.15 & 0.152 & 12.72 & 4.27 \\ \hline

    \end{tabular}%
    }
\end{table*}

\begin{table*}[!ht]
    \centering
    \caption{Evaluation of Robustness Metrics on Three Large Language models for Code Using the Humaneval Dataset Perturbed in the Code. An upward arrow ($\uparrow$) indicates that a higher value corresponds to better robustness, while a downward arrow ($\downarrow$) indicates the opposite. Codegen-2B-mono shows the best performance}
    \label{tab:results_RQ1_2}
    \resizebox{\textwidth}{!}{
    \begin{tabular}{ll|lll|lll|lll}
    \hline
       \textbf{ Perturbations} &~ & \multicolumn{3}{c}{\textbf{codegen-2B-mono}} &\multicolumn{3}{|c|}{\textbf{codegen-2B-multi}} &\multicolumn{3}{c}{\textbf{incoder-1B}}   \\ \hline
        \textbf{Level} & \textbf{Method} & \textbf{passatk}~\textcolor{blue}{$\uparrow$} &\textbf{drop(\%)}~\textcolor{red}{$\downarrow$} & \textbf{rel.(\%)}~\textcolor{red}{$\downarrow$} &\textbf{passatk}~\textcolor{blue}{$\uparrow$} & \textbf{drop(\%)}~\textcolor{red}{$\downarrow$} & \textbf{rel.(\%)}~\textcolor{red}{$\downarrow$} &\textbf{passatk}~\textcolor{blue}{$\uparrow$}& \textbf{drop(\%)}~\textcolor{red}{$\downarrow$} & \textbf{rel.(\%)}~\textcolor{red}{$\downarrow$}  \\ \hline
        \multirow{4}{*}{Character \textbf{(C)}}
        ~ & C1 & 0.250 & 0.00 & 2.44 & 0.116 & 17.39 & 2.44  & 0.0793 & 23.53 & 2.44 \\ \cline{2-11}
        ~ & C2 & 0.220 & 12.20 & 9.15  & 0.122 & 13.04 & 5.49  & 0.0976 & 5.88 & 4.27  \\ \cline{2-11}
        ~ & C3 & 0.409 & 0.00 & 2.44  & 0.281 & 6.12 & 3.05 &  0.1463 & 22.58 & 7.93  \\ \cline{2-11}
        ~ & C4 & 0.396 & 2.99 & 3.66 &  0.287 & 4.08 & 2.44 & 0.1646 & 12.90 & 3.66  \\ \cline{2-11}
        ~ & Median & 0.323 & 1.49 & 3.05 &  0.201 & 9.58 & 2.74 &  0.1220 & 17.74 & 3.96  \\ \hline
        \multirow{6}{*}{Word \textbf{(W)}} 
        ~ & W1 & 0.250 & 0.00 & 2.44 &  0.159 & -13.04 & 1.83 &  0.098 & 5.88 & 3.05 \\ \cline{2-11}
        ~ & W2 & 0.207 & 17.07 & 5.49 &  0.116 & 17.39 & 4.88 &  0.098 & 5.88 & 3.05\\ \cline{2-11}
        ~ & W3 & 0.232 & 7.32 & 6.71 &  0.116 & 17.39 & 4.88 &  0.085 & 17.65 & 3.05 \\ \cline{2-11}
        ~ & W4 & 0.402 & 1.49 & 5.49 &  0.256 & 14.29 & 7.93 &  0.140 & 25.81 & 8.54  \\ \cline{2-11}
        ~ & W5 & 0.378 & 7.46 & 4.27 &  0.244 & 18.37 & 9.15 &  0.128 & 32.26 & 8.54  \\ \cline{2-11}
        ~ & W6 & 0.390 & 4.48 & 5.49 &  0.232 & 22.45 & 9.15 &  0.165 & 12.90 & 6.10  \\ \cline{2-11}
        ~ & Median & 0.314 & 5.90 & 5.49 &  0.195 & 17.39 & 6.40 &  0.113 & 15.28 & 4.57  \\ \hline
        \multirow{6}{*}{Statement \textbf{(S)}} 
        ~ & S1 & 0.256 & -2.44 & 3.05  & 0.140 & 0.00 & 2.44  & 0.116 & -11.76 & 2.44  \\ \cline{2-11}
        ~ & S2 & 0.262 & 35.82 & 18.29 &  0.159 & 46.94 & 15.24 &  0.104 & 45.16 & 9.76 \\ \cline{2-11}
        ~ & S3 & 0.396 & 2.99 & 3.66 &  0.232 & 22.45 & 6.71 &  0.152 & 19.35 & 6.10  \\ \cline{2-11}
        ~ & S4 & 0.409 & 0.00 & 4.88 &  0.244 & 18.37 & 6.71 &  0.177 & 6.45 & 3.66 \\ \cline{2-11}
        ~ & S5 & 0.415 & -1.49 & 5.49 &  0.281 & 6.12 & 5.49 &  0.171 & 9.68 & 3.05 \\ \cline{2-11}
        ~ & S6 & 0.427 & -4.48 & 5.49 &  0.274 & 8.16 & 7.32 &  0.165 & 12.90 & 4.88  \\ \cline{2-11}
        ~ & Median & 0.402 & -0.75 & 5.18 &  0.238 & 13.27 & 6.71 &  0.159 & 11.29 & 4.27  \\ \hline
    \end{tabular}%
    }
\end{table*}

\begin{table*}[!ht]
    \caption{Evaluation of Robustness Metrics on Three Code-Task Models Using the MBPP Dataset Perturbed in the Code. An upward arrow ($\uparrow$) indicates that a higher value corresponds to better robustness, while a downward arrow ($\downarrow$) indicates the opposite. Meta-Llama-3-8B-Instruct shows the best performance}
    \label{tab:results_RQ1_3}
    \resizebox{\textwidth}{!}{
    \begin{tabular}{ll|lll|lll|lll}
     \hline
       \textbf{Perturbations} & ~ & \multicolumn{3}{c}{\textbf{Meta-Llama-3-8B-Instruct}} & \multicolumn{3}{|c|}{\textbf{gpt-j-6b}} &\multicolumn{3}{c}{\textbf{CodeLlama-7b-hf}} \\ \hline
        \textbf{Level} & \textbf{Method} & \textbf{passatk}~\textcolor{blue}{$\uparrow$} &\textbf{drop(\%)}~\textcolor{red}{$\downarrow$} & \textbf{rel.(\%)}~\textcolor{red}{$\downarrow$} &\textbf{passatk}~\textcolor{blue}{$\uparrow$} & \textbf{drop(\%)}~\textcolor{red}{$\downarrow$} & \textbf{rel.(\%)}~\textcolor{red}{$\downarrow$} &\textbf{passatk}~\textcolor{blue}{$\uparrow$}& \textbf{drop(\%)}~\textcolor{red}{$\downarrow$} & \textbf{rel.(\%)}~\textcolor{red}{$\downarrow$}  \\ \hline
        
        \multirow{4}{*}{Character (\textbf{C})}
        ~ & C1 & 0.593 & 1.37 & 3.90 & 0.120 & 11.36 & 3.39 & 0.059 & 9.52 & 1.23 \\ \cline{2-11}
        ~ & C2 & 0.602 & 0.00 & 9.03 & 0.109 & 19.70 & 9.65 & 0.073 & -12.70 & 4.72 \\ \cline{2-11}
        ~ & C3 & 0.519 & 15.27 & 18.79 & 0.176 & 0.58 & 8.32 & 0.502 & -137.38 & 35.42 \\ \cline{2-11}
        ~ & C4 & 0.584 & 4.53 & 8.11 & 0.160 & 9.30 & 4.72 & 0.220 & -3.88 & 4.11 \\ \cline{2-11}
        ~ & Median & 0.589 & 2.95 & 8.57 & 0.140 & 10.33 & 6.52 & 0.146 & -8.29 & 4.41 \\ \hline
    
        \multirow{6}{*}{Word (\textbf{W})} 
        ~ & W1 & 0.605 & -0.51 & 5.44 & 0.134 & 1.52 & 3.08 & 0.090 & -39.68 & 4.00 \\ \cline{2-11}
        ~ & W2 & 0.596 & 1.02 & 8.42 & 0.115 & 15.15 & 7.39 & 0.064 & 1.59 & 3.18 \\ \cline{2-11}
        ~ & W3 & 0.600 & 0.17 & 0.10 & 0.135 & 0.76 & 0.10 & 0.065 & 0.00 & 0.00 \\ \cline{2-11}
        ~ & W4 & 0.612 & 0.00 & 13.14 & 0.157 & 11.05 & 8.73 & 0.153 & 27.67 & 9.14 \\ \cline{2-11}
        ~ & W5 & 0.607 & 0.84 & 14.48 & 0.127 & 27.91 & 11.09 & 0.155 & 26.70 & 9.34 \\ \cline{2-11}
        ~ & W6 & 0.611 & 0.17 & 14.07 & 0.136 & 23.26 & 10.88 & 0.158 & 25.24 & 8.62 \\ \cline{2-11}
        ~ & Median & 0.606 & 0.17 & 10.78 & 0.134 & 13.10 & 8.06 & 0.122 & 13.42 & 6.31 \\ \hline
    
        \multirow{6}{*}{Statement (\textbf{S})} 
        ~& S1 & 0.598 & 0.68 & 3.70 & 0.127 & 6.06 & 2.87 & 0.072 & -11.11 & 1.75 \\ \cline{2-11}
        ~ & S2 & 0.435 & 28.86 & 30.80 & 0.072 & 59.30 & 15.20 & 0.055 & 73.79 & 17.25 \\ \cline{2-11}
        ~ & S3 & 0.610 & 0.34 & 12.11 & 0.150 & 15.12 & 9.45 & 0.096 & 54.85 & 12.83 \\ \cline{2-11}
        ~ & S4 & 0.628 & -2.68 & 11.09 & 0.160 & 9.30 & 9.45 & 0.157 & 25.73 & 8.11 \\ \cline{2-11}
        ~ & S5 & 0.611 & 0.17 & 7.91 & 0.146 & 17.44 & 5.75 & 0.209 & 0.97 & 5.75 \\ \cline{2-11}
        ~ & S6 & 0.599 & 2.18 & 8.93 & 0.149 & 15.70 & 7.70 & 0.168 & 20.39 & 7.60 \\ \cline{2-11}
        ~ & Median & 0.604 & 0.51 & 10.01 & 0.147 & 15.41 & 8.57 & 0.126 & 23.06 & 7.85 \\ \hline

    \end{tabular}%
    }
\end{table*}

\begin{table*}[!ht]
    \centering
    \caption{Evaluation of Robustness Metrics on Three Large Language models for Code Using the MBPP Dataset Perturbed in the Code. An upward arrow ($\uparrow$) indicates that a higher value corresponds to better robustness, while a downward arrow ($\downarrow$) indicates the opposite. Codegen-2B-mono shows the best performance}
    \label{tab:results_RQ1_4}
     \resizebox{\textwidth}{!}{
     \begin{tabular}{ll|lll|lll|lll}
    \hline
       \textbf{ Perturbations} &~ & \multicolumn{3}{c}{\textbf{codegen-2B-mono}} &\multicolumn{3}{|c|}{\textbf{codegen-2B-multi}} &\multicolumn{3}{c}{\textbf{incoder-1B}}   \\ \hline
        \textbf{Level} & \textbf{Method} & \textbf{passatk}~\textcolor{blue}{$\uparrow$} &\textbf{drop(\%)}~\textcolor{red}{$\downarrow$} & \textbf{rel.(\%)}~\textcolor{red}{$\downarrow$} &\textbf{passatk}~\textcolor{blue}{$\uparrow$} & \textbf{drop(\%)}~\textcolor{red}{$\downarrow$} & \textbf{rel.(\%)}~\textcolor{red}{$\downarrow$} &\textbf{passatk}~\textcolor{blue}{$\uparrow$}& \textbf{drop(\%)}~\textcolor{red}{$\downarrow$} & \textbf{rel.(\%)}~\textcolor{red}{$\downarrow$}  \\ \hline
        \multirow{4}{*}{Character \textbf{(C)}}
        ~ & C1 & 0.309 & 2.59 & 4.41 & 0.185 & 1.08 & 4.52 & 0.116 & 9.60 & 3.18 \\ \cline{2-11}
        ~ & C2 & 0.295 & 7.12 & 9.55 & 0.179 & 6.45 & 10.88 & 0.093 & 27.20 & 8.32  \\ \cline{2-11}
        ~ & C3 & 0.415 & -3.03 & 3.66 & 0.305 & -4.17 & 4.88 & 0.146 & 22.58 & 7.93  \\ \cline{2-11}
        ~ & C4 & 0.384 & 4.55 & 3.05 & 0.274 & 6.25 & 4.27 & 0.171 & 9.68 & 3.05  \\ \cline{2-11}
        ~ & Median & 0.347 & 3.57 & 4.04 & 0.230 & 3.67 & 4.70 & 0.131 & 16.13 & 5.56  \\ \hline
        \multirow{6}{*}{Word \textbf{(W)}} 
        ~ & W1 & 0.316 & 0.32 & 5.44 & 0.196 & -2.69 & 5.44 & 0.116 & 9.60 & 5.75 \\ \cline{2-11}
        ~ & W2 & 0.312 & 1.62 & 7.19 & 0.185 & 3.23 & 8.62 & 0.110 & 14.40 & 6.67\\ \cline{2-11}
        ~ & W3 & 0.316 & 0.32 & 3.08 & 0.186 & 2.69 & 7.49 & 0.105 & 18.40 & 7.70 \\ \cline{2-11}
        ~ & W4 & 0.428 & 4.79 & 15.30 & 0.263 & 7.91 & 13.96 & 0.194 & 11.27 & 13.14  \\ \cline{2-11}
        ~ & W5 & 0.417 & 7.31 & 16.63 & 0.240 & 15.83 & 15.61 & 0.171 & 21.60 & 14.17  \\ \cline{2-11}
        ~ & W6 & 0.355 & 21.00 & 22.90 & 0.191 & 33.09 & 22.90 & 0.114 & 47.89 & 23.82 \\ \cline{2-11}
        ~ & Median & 0.336 & 3.21 & 11.25 & 0.194 & 5.57 & 11.29 & 0.115 & 16.40 & 10.42  \\ \hline
        \multirow{6}{*}{Statement \textbf{(S)}} 
        ~ & S1 &  0.318 & -0.32 & 3.08 & 0.187 & 2.15 & 4.31 & 0.128 & 0.00 & 2.46  \\ \cline{2-11}
        ~ & S2 & 0.043 & 90.41 & 52.05 & 0.020 & 93.17 & 37.99 & 0.015 & 92.96 & 29.26 \\ \cline{2-11}
        ~ & S3 & 0.432 & 3.88 & 13.66 & 0.259 & 9.35 & 12.94 & 0.182 & 16.90 & 12.73  \\ \cline{2-11}
        ~ & S4 & 0.450 & 0.00 & 13.24 & 0.275 & 3.60 & 11.81 & 0.225 & -2.82 & 12.32 \\ \cline{2-11}
        ~ &S5 & 0.360 & 10.61 & 12.20 & 0.220 & 25.00 & 10.98 & 0.152 & 19.35 & 7.32 \\ \cline{2-11}
        ~ & S6 & 0.409 & -1.52 & 6.71 & 0.262 & 10.42 & 7.93 & 0.165 & 12.90 & 4.88  \\ \cline{2-11}
        ~ & Median & 0.385 & 1.94 & 12.72 & 0.240 & 9.89 & 11.40 & 0.159 & 14.90 & 9.82  \\ \hline
    \end{tabular}*
    }
\end{table*}

\begin{table}[!ht]
    \centering
    \caption{Wilcoxon signed-ranked test($\alpha$ = 0.05) for RQ1. We applied the Wilcoxon signed-rank test to compare model performance under different perturbation conditions and used Bonferroni correction to adjust for multiple comparisons. The majority of p-values are below 0.05 and 87\% of comparisons show large effect sizes, reinforcing the practical impact of these differences.}
    \label{tab:wilcoxon_RQ1}
    \resizebox{\textwidth}{!}{
    \begin{tabular}{|l|l|l|l|l|l|l|l|l|}
    \hline
        Wilcoxon Statistic & p-value & Bonferroni Correction & Effect Size (r) & ~ & Wilcoxon Statistic & p-value & Bonferroni Correction & Effect Size (r)  \\ \hline
        
        0 & 0.000195066 & 0.00292599 & 0.87988269 & ~ & 0 & 0.000436809 & 0.006552135 & 0.87988269  \\ \hline
        1 & 0.000232035 & 0.003480525 & 0.865217979 & ~ & 3 & 0.000152588 & 0.00228882 & 0.835888556  \\ \hline
        0 & 0.000196093 & 0.002941395 & 0.87988269 & ~ & 1 & 6.10E-05 & 0.000915 & 0.865217979  \\ \hline
        0 & 0.000194724 & 0.00292086 & 0.87988269 & ~ & 0 & 3.05E-05 & 0.0004575 & 0.87988269  \\ \hline
        0 & 0.000192345 & 0.002885175 & 0.87988269 & ~ & 0 & 3.05E-05 & 0.0004575 & 0.87988269  \\ \hline
        18 & 0.003274769 & 0.049121535 & 0.615917883 & ~ & 19 & 0.009185791 & 0.137786865 & 0.601253172  \\ \hline
        1 & 0.000350861 & 0.005262915 & 0.865217979 & ~ & 2 & 9.16E-05 & 0.001374 & 0.850553267  \\ \hline
        0 & 0.00019575 & 0.00293625 & 0.87988269 & ~ & 0 & 3.05E-05 & 0.0004575 & 0.87988269  \\ \hline
        0 & 0.00019575 & 0.00293625 & 0.87988269 & ~ & 0 & 3.05E-05 & 0.0004575 & 0.87988269  \\ \hline
        19 & 0.003770074 & 0.05655111 & 0.601253172 & ~ & 57 & 0.596588135 & 8.948822025 & 0.043994135  \\ \hline
        68 & 0.445657924 & 6.68486886 & 0.117317692 & ~ & 53 & 0.463745117 & 6.956176755 & 0.102652981  \\ \hline
        0 & 0.000194724 & 0.00292086 & 0.87988269 & ~ & 0 & 3.05E-05 & 0.0004575 & 0.87988269  \\ \hline
        6 & 0.000530787 & 0.007961805 & 0.791894421 & ~ & 64 & 0.86026001 & 12.90390015 & 0.058658846  \\ \hline
        0 & 0.00019575 & 0.00293625 & 0.87988269 & ~ & 0 & 3.05E-05 & 0.0004575 & 0.87988269  \\ \hline
        0 & 0.000188984 & 0.00283476 & 0.87988269 & ~ & 0 & 3.05E-05 & 0.0004575 & 0.87988269 \\ \hline
    \end{tabular}
    }
\end{table}

\noindent\textbf{Results and Analysis:}
Tables~\ref{tab:results_RQ1_1} and ~\ref{tab:results_RQ1_2} present the experimental results of the models against the HumanEval dataset, subject to the aforementioned three levels of perturbations.
The analysis of Large Language models for Code at different levels of perturbation (character, word, and statement) provides valuable insights into model robustness. 
Under the statement-level perturbation, the model Llama-3-8B-Instruct, for example, recorded a median value of RP 0.652 in the HumanEval dataset, higher than in word-level (0.616) and character-level (0.616). This implies that Llama-3-8B-Instruct has the best robustness under statement-level perturbation. Given that word-level and character-level perturbation had the same RP value, we compared their RD median value. Under word-level perturbation, the median RD value is 1.83\%, which is higher than its character-level value (0.92\%), which means this models have better robustness under character-level than word-level in terms of the RD metric. \\
Across the majority of (five out of six) models, we observe a consistent trend of enhanced robustness to statement-level perturbations within the HumanEval dataset. These models also demonstrate moderate robustness to character-level perturbations and less robustness to word-level perturbations. Notably, the Meta-Llama-3-8B-Instruct models stand out in terms of robustness, according to the reported RP median values under character-level, word-level, and statement-level perturbations (0.616, 0.616, and 0.652, respectively). In contrast, the CodeLlama-7b-hf model exhibits the least robustness, with median RP values under character-level, word-level, and statement-level perturbations of 0.113, 0.110, and 0.152, respectively. 
Tables~\ref{tab:results_RQ1_3} and~\ref{tab:results_RQ1_4} present the experimental results on the MBPP dataset under three levels of perturbations. 
In Table~\ref{tab:results_RQ1_3}, the median \textbf{RP} values of Meta-Llama-3-8B-Instruct across the different perturbation levels are 0.589, 0.606, and 0.604, respectively. Given the similarity between the word-level and statement-level \textbf{RP} values, we compared the median \textbf{RD} values, which are 0.17\% and 0.51\%. These results indicate that Meta-Llama-3-8B-Instruct demonstrates the highest robustness under word-level perturbations, followed by statement-level perturbations, with the least robustness under character-level perturbations. However, when analyzing the results for GPT-J-6B, the \textbf{RP} values of 0.140, 0.134, and 0.147 suggest that GPT-J-6B shows the greatest robustness under statement-level perturbations, followed by character-level perturbations, with the lowest performance under word-level perturbations. \\
Table~\ref{tab:results_RQ1_4} reveals a consistent robustness pattern for all three models, mirroring the trends observed on the HumanEval dataset, as seen in Table~\ref{tab:results_RQ1_2}. Similarly, Meta-Llama-3-8B-Instruct exhibits the highest robustness, while CodeLlama-7b-hf demonstrates the weakest performance on the MBPP dataset. 
An in-depth analysis of the CodeGen family shows that an increase in parameters (e.g. codegen-2B-mono, codegen-350M-mono) is associated with a noticeable improvement in robustness pass (RP). However, when comparing mono and multi language models, it is observed that robustness drop (RD) and robustness relative (RR) metrics display significant variability. 

Mono language models demonstrate a distinct decrease in RD without significant changes in RR, indicating enhanced robustness across all perturbation levels. On the other hand, in multi-language models, RP decreases significantly while RD and RR increase. This comparative analysis suggests that mono language models exhibit better robustness compared to multi language models. All models perform best with \textbf{statement-level perturbations}. However, the DeadCodeInserter perturbation has a significant influence on models due to increased complexity and context confusion. Deadcode, which refers to irrelevant or unnecessary lines of code, adds unnecessary noise to the codebase, increasing its complexity and making it harder for the model to distinguish between relevant and irrelevant parts of the code, thus leading to potential misinterpretations. Models rely on contextual information to make accurate predictions and code completions. Deadcode can create context overload, presenting the model with too much irrelevant information and diluting the useful context needed for correct code generation. Specifically, for the CodeLlama-7b-hf model, the tab\_indent perturbation has a particularly large impact. This reveals that the CodeLlama-7b-hf model is sensitive to tabs and spaces. Implementing standard unification practices can better guarantee the model's performance. Thus, models exhibit the highest robustness to statement-level perturbations because they can utilize the broader context and structure of the code to infer the intended functionality. They are moderately resilient to character-level perturbations due to their ability to detect and correct minor errors based on patterns and context. However, they are most susceptible to word-level perturbations because these can significantly alter the meaning and relationships within the code, making it difficult for the model to interpret and generate the correct output accurately.

\newpage
\begin{answer*}{ (RQ1)}{}
Perturbations of code at the character, word, and statement levels impact the functional correctness of LLM4Code differently. The results indicate that models are most resilient to sentence-level perturbations, followed by character-level perturbations, with word-level perturbations being the most detrimental to performance. Additionally, the robustness and effectiveness of models are significantly influenced by the specific characteristics and complexities of the datasets used for evaluation. Also, we find that mono-language models often perform better than multi-language models.
\end{answer*}


\noindent \textbf{\rqtwo}\\
\noindent\minor{\textbf{Scope:} Perturbations are applied to docstrings or task descriptions, typically found at the beginning of the prompt.}\\
\noindent\textbf{Objective:} 
This research question focuses on determining the impact of perturbations applied to the task description (\textit{the natural language perspective}) at different levels—character-level, word-level, and sentence-level—on the functional correctness of the code generated by models. The objective is to assess how variations in the clarity and accuracy of task descriptions influence the ability of Large Language models for Code to produce functionally correct and semantically accurate code. By analyzing the code output in response to progressively complex perturbations in the task descriptions, this study aims to uncover the extent to which the fidelity of natural language instructions affects the models' performance. The findings will offer insights into the robustness of Large Language models for Code to inaccuracies or manipulations in natural language inputs, highlighting the importance of robust natural language understanding in the context of automated code generation.

\noindent\textbf{Approach:} 
In the present study, the OpenAttack toolkit is employed to introduce perturbations into the task description component of datasets, notably HumanEval and MBPP. The initial phase of this research involves the extraction of the ``task description'' segment from the foundational dataset, followed by its preservation in a newly formulated dataset file, annotated with the corresponding task\_id. Subsequent to this preparatory step, a variety of adversarial attack models are applied to instigate alterations within the ``task description'' dataset. These modified datasets, containing the perturbed task descriptions, are then integrated back into the original database, replacing the unaltered descriptions. \edit{All perturbations in our study are derived from prior work (ReCode and OpenAttack), ensuring their relevance in robustness testing. While some perturbations may be uncommon in everyday software development, they pose potential security risks, as attackers can intentionally introduce them to manipulate LLM4Code. As adversarial robustness evaluation aims to uncover model vulnerabilities, it is essential to test LLM4Code against both frequent and rare perturbations to ensure resilience against a wide range of threats.}The final stage of the investigation entails the evaluation of models utilizing this revised dataset, from which three principal metrics are derived: RPs@k (Robust Pass@k), RDs@k (Robust Drops@k), and RRs@k (Robust Relatives@k), as delineated in Section~\ref{sec:eval-metrics}. 
In Listings~\ref{lst:butterfingers}~\ref{lst:insertion}~\ref{lst:backtranslation}, we provide examples of different types of perturbations to the original task description shown in Listing~\ref{lst:original}. Listing~\ref{lst:butterfingers} shows a task description, reading from line 3 through line 4, in which \texttt{`from'} and \texttt{`rolling'} are changed to \texttt{`frol'} and \texttt{`rollibg'}. Listing~\ref{lst:insertion} provides another task description. In lines 3 through 4, we introduced synonyms to create variations in the text. For instance, we appended \texttt{beget} to \texttt{generate}, transforming \texttt{generate} into \texttt{generate beget}. Similarly, \texttt{maximum} was changed to \texttt{maximum maximal}, and \texttt{moment} became \texttt{moment here and now}, among other examples. Listing~\ref{lst:backtranslation} demonstrates that the description from lines 3 to 4 was translated into Mandarin and then back into English. \edit{To ensure statistical validity, we applied the Wilcoxon signed-rank test to compare model performance under different perturbation conditions and used Bonferroni correction to adjust for multiple comparisons and control the family-wise error rate. The detailed results of these statistical tests are presented in Section~\ref{sec:discussion}}

\edit{Table~\ref{tab:reclassification_for_RQ2} presents the definitions and classifications of various perturbations provided in ReCode. In addition, five other perturbation models with a high attack success rate in OpenAttack are listed in Table~\ref{tab:reclassification_for_RQ2.2}. }

\begin{table*}[!ht]
\centering
\caption{Classifying task description perturbations in ReCode into three classes: Character-level(C), Word-level(W), and Statement-Level(S)}
\label{tab:reclassification_for_RQ2}
\resizebox{\linewidth}{!}{
\begin{threeparttable}
    \begin{tabular}{l|l|l} 
\hline
\hline
\textbf{Perturbation Method}          & \textbf{Definition}   & \textbf{Classification } \\
\hline
\hline
ButterFingersPerturbation \textbf{(C1)}   & Involves random character changes or typos, affecting characters.            & Character-Level\\
\hline
ChangeCharCase\textbf{(C2)}              & Changing the case of characters.                                               & Character-Level\\ 
\hline
SwapCharactersPerturbation \textbf{(C3)}  & Swapping characters within words.                                               & Character-Level\\
\hline
WhitespacePerturbation \textbf{(C4)}      & Involves changes in the use of spaces, tabs, or other whitespace characters.      & Character-Level\\ 
\hline
SynonymInsertion \textbf{(W1)}            & Inserting synonyms, which affects whole words.                                    & Word-Level \\ 
\hline
SynonymSubstitution \textbf{(W2)}         & Substituting words with their synonyms.                                          & Word-Level \\
\hline
EnglishInflectionalVariation \textbf{(W3)}    & Altering the inflection of words which can change the meaning or structure of sentences. & Word-Level\\ 
\hline
BackTranslation \textbf{(S1) }            & Translating text to another language and back, affecting entire sentences.     & Sentence-Level\\
\hline
TenseTransformationPast \textbf{(S2)}     & Changing the tense of sentences to the past tense.                                & Sentence-Level\\ 
\hline
TenseTransformationFuture \textbf{(S3)}   & Changing the tense of sentences to the future tense.                              & Sentence-Level\\
\hline
new\_line\_afterdoc \textbf{(S4)}           & Inserting new lines after documentation lines, affecting the documentation structure.& Statement-Level\\
\hline
\end{tabular}
\end{threeparttable}
}
\end{table*}


\begin{table*}[!ht]
\centering
\caption{Classifying task description perturbations(high successful attack rate) in OpenAttack toolbox into three classes: Character-level(C), Word-level(W), and Statement-Level(S)}
\label{tab:reclassification_for_RQ2.2}
\resizebox{\linewidth}{!}{
\begin{threeparttable}
    \begin{tabular}{l|l|l} 
\hline
\hline
\textbf{Perturbation Method}          & \textbf{main idea}   & \textbf{Classification }\\
\hline
\hline
PWWS (Probability Weighted Word Saliency) &Greedy word substitution &Word-Level\\
\hline
SememePSO &Particle Swarm Optimization-based word substitution &word-Level\\
\hline
FD (Feature Disruption) &Gradient-based word substitution &word-Level\\
\hline
TextBugger &Greedy word substitution and character manipulation &Word+Character-Level\\
\hline
HotFlip &Gradient-based word or character substitution &Word+Character-Level\\
\hline
DeepWordBug &Greedy character manipulation & Character-Level\\
\hline
\end{tabular}
\end{threeparttable}
}
\end{table*}
\noindent\begin{minipage}{.47\textwidth}
\begin{lstlisting}[escapechar=!, language=Python, label=lst:butterfingers, caption=``ButterFingersPerturbation''; Character-level perturbations. Task description lines 3-4: indicates that ``from" and ``rolling'' are changed to ``Frol'' and "rollibg''.]
def rolling_max(numbers):
    """
    !\colorbox{yellow}{Frol}! a given list of integers, generate a list of !\colorbox{yellow}{rollibg}! maxmmum element found until given moment in the sequenct.
    >>> rolling_max([1, 2, 3, 2, 3, 4, 2])
    [1, 2, 3, 3, 3, 4, 4]
    """
    running_max = None
    result = []

    for n in numbers:
        if running_max is None:
            running_max = n
        else:
            running_max = max(running_max, n)
        result.append(running_max)
    return result
\end{lstlisting}
\end{minipage}\hfill
\begin{minipage}{.47\textwidth}
\begin{lstlisting}[escapechar=!, language=Python, label=lst:insertion, caption=``SynonymInsertion''; Word-level perturbations. Task description shown in lines 3-4 indicates that synonyms are added after certain words: ``beget'' after ``generate'']
def rolling_max(numbers):
    """
    From a given list of integers, generate !\colorbox{yellow}{beget}! a list of rolling maximum maximal element found determine until given moment !\colorbox{yellow}{here and now}! in the sequence.
    >>> rolling_max([1, 2, 3, 2, 3, 4, 2])
    [1, 2, 3, 3, 3, 4, 4]
    """
    running_max = None
    result = []

    for n in numbers:
        if running_max is None:
            running_max = n
        else:
            running_max = max(running_max, n)
        result.append(running_max)
    return result
\end{lstlisting}
\end{minipage}\hfill

\begin{minipage}{0.95\textwidth}
\begin{lstlisting}[escapechar=!, language=Python, label=lst:backtranslation, caption=An example of sentence-level perturbations; BackTranslation. This description line 3 to line 4 was translated into German and then back into English.]
def rolling_max(numbers):
    """
    !\colorbox{yellow}{Generate from a given list of integers a list of rotating maximum elements found up to a certain moment in the sequence.}!
    >>> rolling_max([1, 2, 3, 2, 3, 4, 2])
    [1, 2, 3, 3, 3, 4, 4]
    """
    running_max = None
    result = []

    for n in numbers:
        if running_max is None:
            running_max = n
        else:
            running_max = max(running_max, n)
        result.append(running_max)
    return result
\end{lstlisting}
\end{minipage}\hfill

\begin{table*}[!ht]
    \centering
    \caption{
    Evaluation of robustness metrics in three code-task models using the Humaneval dataset perturbed in the task description. An upward arrow ($\uparrow$) indicates that a higher value corresponds to better robustness, while a downward arrow ($\downarrow$) indicates the opposite. Meta-Llama-3-8B-Instruct shows the best performance in terms of Robustness.
    }
    \label{tab:results_RQ2_1}
     \resizebox{\textwidth}{!}{
     \begin{tabular}{ll|lll|lll|lll}
     \hline
       \textbf{Perturbations} & ~ & \multicolumn{3}{c}{\textbf{Meta-Llama-3-8B-Instruct}} &  \multicolumn{3}{|c|}{\textbf{gpt-j-6b}} &\multicolumn{3}{c}{\textbf{CodeLlama-7b-hf}}\\ \hline
       Level & Method & passatk~\textcolor{blue}{$\uparrow$} &drop(\%)~\textcolor{red}{$\downarrow$} & rel.(\%)~\textcolor{red}{$\downarrow$} &passatk~\textcolor{blue}{$\uparrow$} & drop(\%)~\textcolor{red}{$\downarrow$} & rel.(\%)~\textcolor{red}{$\downarrow$} &passatk~\textcolor{blue}{$\uparrow$}& drop(\%)~\textcolor{red}{$\downarrow$} & rel.(\%)~\textcolor{red}{$\downarrow$} \\ \hline
        \multirow{2}{*}{Word/Char} 
        ~ & TextBugger & 0.57 & 1.05 & 1.83 & 0.10 & 15.79 & 1.83 & 0.04 & 22.22 & 1.22 \\ \cline{2-11}
        ~ & HotFlip & 0.58 & 0.00 & 1.22 & 0.10 & 15.79 & 3.05 & 0.04 & 22.22 & 1.22 \\ \cline{2-11}
        ~ & Median & 0.58 & 0.53 & 1.53 & 0.10 & 15.79 & 2.44 & 0.04 & 22.22 & 1.22 \\ \hline
        \multirow{5}{*}{Character} 
        ~ & DeepWordBug & 0.56 & 3.16 & 1.83 & 0.09 & 26.32 & 4.27 & 0.04 & 22.22 & 1.22 \\ \cline{2-11}
        ~ & C1 & 0.55 & 5.26 & 11.59 & 0.10 & 10.53 & 3.66 & 0.02 & 55.56 & 4.27 \\ \cline{2-11}
        ~ & C2 & 0.48 & 16.84 & 13.41 & 0.07 & 36.84 & 5.49 & 0.02 & 55.56 & 5.49 \\ \cline{2-11}
        ~ & C3 & 0.58 & 0.00 & 7.32 & 0.09 & 21.05 & 3.66 & 0.02 & 55.56 & 4.27 \\ \cline{2-11}
        ~ & C4 & 0.54 & 7.37 & 10.37 & 0.04 & 22.22 & 3.66 & 0.10 & 15.79 & 1.83 \\ \cline{2-11}
        ~ & Median & 0.55 & 5.26 & 10.37 & 0.09 & 22.22 & 3.66 & 0.02 & 55.56 & 4.27 \\ \hline
        \multirow{6}{*}{Word} 
        ~ & PWWS & 0.57 & 1.05 & 1.83 & 0.09 & 26.32 & 6.71 & 0.04 & 22.22 & 1.22 \\ \cline{2-11}
        ~ & PSO & 0.58 & 0.00 & 1.22 & 0.08 & 31.58 & 3.66 & 0.04 & 22.22 & 1.22 \\ \cline{2-11}
        ~ & FD & 0.58 & 0.00 & 1.22 & 0.09 & 26.32 & 4.27 & 0.04 & 22.22 & 1.22 \\ \cline{2-11}
        ~ & W1 & 0.51 & 12.63 & 10.98 & 0.12 & -5.26 & 1.83 & 0.06 & -11.11 & 4.27 \\ \cline{2-11}
        ~ & W2 & 0.48 & 16.84 & 15.85 & 0.09 & 21.05 & 2.44 & 0.02 & 66.67 & 6.10 \\ \cline{2-11}
        ~ & W3 & 0.52 & 9.47 & 10.37 & 0.10 & 10.53 & 3.66 & 0.01 & 88.89 & 6.10 \\ \cline{2-11}
        ~ & Median & 0.55 & 5.26 & 6.1 & 0.09 & 23.69 & 3.66 & 0.04 & 22.22 & 2.75 \\ \hline
        \multirow{4}{*}{Sentence} 
        ~ & S1 & 0.49 & 14.74 & 9.76 & 0.10 & 15.79 & 3.05 & 0.06 & -11.11 & 1.83 \\ \cline{2-11}
        ~ & S2 & 0.57 & 1.05 & 7.93 & 0.27 & 4.26 & 7.32 & 0.03 & 44.44 & 3.66 \\ \cline{2-11}
        ~ & S3 & 0.56 & 3.16 & 11.59 & 0.25 & 12.77 & 7.32 & 0.03 & 44.44 & 3.66 \\ \cline{2-11}
        ~ & S4 & 0.69 & -3.67 & 4.88 & 0.52 & 6.52 & 4.88 & 0.21 & -9.68 & 3.05 \\ \cline{2-11}
        ~ & Median & 0.57 & 2.11 & 8.85 & 0.26 & 9.65 & 6.1 & 0.05 & 16.67 & 3.36 \\ \hline

    \end{tabular}*
    }
\end{table*}

\begin{table*}[!ht]
    \centering
    \caption{Evaluation of Robustness Metrics on Three Large Language models for Code Using the Humaneval Dataset Perturbed in the task description. An upward arrow ($\uparrow$) indicates that a higher value corresponds to better robustness, while a downward arrow ($\downarrow$) indicates the opposite. Codegen-2B-mono shows the best performance in terms of ROBUSTNESS.}
    \label{tab:results_RQ2_2}
     \resizebox{\textwidth}{!}{
     \begin{tabular}{ll|lll|lll|lll}
    \hline
        \textbf{ Perturbations} &~ & \multicolumn{3}{c}{\textbf{codegen-2B-mono}} &\multicolumn{3}{|c|}{\textbf{codegen-2B-multi}} &\multicolumn{3}{c}{\textbf{incoder-1B}}    \\ \hline
        \textbf{Level} & \textbf{Method} & \textbf{passatk}~\textcolor{blue}{$\uparrow$} &\textbf{drop(\%)}~\textcolor{red}{$\downarrow$} & \textbf{rel.(\%)}~\textcolor{red}{$\downarrow$} &\textbf{passatk}~\textcolor{blue}{$\uparrow$} & \textbf{drop(\%)}~\textcolor{red}{$\downarrow$} & \textbf{rel.(\%)}~\textcolor{red}{$\downarrow$} &\textbf{passatk}~\textcolor{blue}{$\uparrow$}& \textbf{drop(\%)}~\textcolor{red}{$\downarrow$} & \textbf{rel.(\%)}~\textcolor{red}{$\downarrow$}  \\ \hline
        \multirow{2}{*}{Word/Char} & TextBugger & 0.23 & 43.28 & 21.34 &  0.12 & 59.18 & 21.34 &  0.10 & 45.16 & 12.20  \\ \cline{2-11}
        ~ & HotFlip & 0.38 & 7.46 & 4.27 &  0.24 & 18.37 & 9.15 &  0.13 & 32.26 & 8.54  \\ \cline{2-11}
        ~ & Median & 0.30 & 25.37 & 12.80 &  0.18 & 38.78 & 15.24 &  0.12 & 38.71 & 10.37  \\ \hline
        \multirow{5}{*}{Character} & DeepWordBug & 0.23 & 44.78 & 20.73 & 0.11 & 63.27 & 20.12 & 0.07 & 61.29 & 15.24  \\ \cline{2-11}
        ~ & C1 & 0.23 & 9.76 & 4.88 & 0.14 & 0.00 & 2.44 & 0.09 & 17.65 & 3.05  \\ \cline{2-11}
        ~ & C2 & 0.18 & 26.83 & 9.15 & 0.13 & 4.35 & 4.27 & 0.10 & 0.00 & 6.10  \\ \cline{2-11}
        ~ & C3 & 0.24 & 4.88 & 6.10 & 0.15 & -8.70 & 2.44 &  0.10 & 0.00 & 3.66  \\ \cline{2-11}
        ~ & C4 & 0.22 & 12.20 & 5.49 & 0.13 & 4.35 & 6.71 & 0.10 & 0.00 & 3.66  \\ \cline{2-11}
        ~ & Median & 0.23 & 12.20 & 6.10 & 0.13 & 4.35 & 3.35 & 0.10 & 0.00 & 3.66  \\ \hline
        \multirow{6}{*}{Word} & PWWS & 0.21 & 49.25 & 22.56 &  0.11 & 63.27 & 21.34 &  0.08 & 58.06 & 13.41  \\ \cline{2-11}
        ~ & PSO & 0.23 & 44.78 & 20.73 & 0.12 & 61.22 & 20.73 &  0.10 & 45.16 & 13.41  \\ \cline{2-11}
        ~ & FD & 0.23 & 43.28 & 21.34 & 0.13 & 57.14 & 20.73 &  0.10 & 45.16 & 13.41  \\ \cline{2-11}
        ~ & W1 & 0.24 & 4.88 & 4.88 &  0.13 & 8.70 & 1.22 &  0.09 & 11.76 & 1.22  \\ \cline{2-11}
        ~ & W2 & 0.23 & 7.32 & 5.49 &  0.13 & 4.35 & 3.05 &  0.09 & 11.76 & 3.66  \\ \cline{2-11}
        ~ & W3 & 0.25 & 0.00 & 6.10 & 0.13 & 8.70 & 3.66 & 0.09 & 11.76 & 2.44  \\ \cline{2-11}
        ~ & Median & 0.23 & 24.08 & 13.42 & 0.13 & 32.92 & 20.73 & 0.09 & 28.46 & 8.54  \\ \hline
        \multirow{4}{*}{Sentence} & S1 & 0.25 & 0.00 & 1.22 & 0.14 & 0.00 & 1.22 & 0.11 & -5.88 & 1.83  \\ \cline{2-11}
        ~ & S2 & 0.41 & -1.49 & 3.05 & 0.29 & 2.04 & 4.27 & 0.18 & 3.23 & 0.61  \\ \cline{2-11}
        ~ & S3 & 0.23 & 9.76 & 2.44 &  0.13 & 8.70 & 4.88 &  0.10 & 0.00 & 2.44  \\ \cline{2-11}
        ~ & S4 & 0.21 & 14.63 & 3.66 &  0.12 & 13.04 & 4.27 & 0.10 & 5.88 & 1.83  \\ \cline{2-11}
        ~ & Median & 0.24 & 4.88 & 2.75 &  0.13 & 5.37 & 4.27 & 0.11 & 1.62 & 1.83 \\ \hline

    \end{tabular}*
    }
\end{table*}

\begin{table*}[!ht]
    \centering
    \caption{
    Evaluation of robustness metrics in three code-task models using the MBPP dataset perturbed in the task description. An upward arrow ($\uparrow$) indicates that a higher value corresponds to better robustness, while a downward arrow ($\downarrow$) indicates the opposite. Meta-Llama-3-8B-Instruct shows the best performance in terms of Robustness.
    }
    \label{tab:results_RQ2_3}
     \resizebox{\textwidth}{!}{
     \begin{tabular}{ll|lll|lll|lll}
     \hline
       \textbf{Perturbations} & ~ & \multicolumn{3}{c}{\textbf{Meta-Llama-3-8B-Instruct}} &  \multicolumn{3}{|c|}{\textbf{gpt-j-6b}} &\multicolumn{3}{c}{\textbf{CodeLlama-7b-hf}}\\ \hline
       Level & Method & passatk~\textcolor{blue}{$\uparrow$} &drop(\%)~\textcolor{red}{$\downarrow$} & rel.(\%)~\textcolor{red}{$\downarrow$} &passatk~\textcolor{blue}{$\uparrow$} & drop(\%)~\textcolor{red}{$\downarrow$} & rel.(\%)~\textcolor{red}{$\downarrow$} &passatk~\textcolor{blue}{$\uparrow$}& drop(\%)~\textcolor{red}{$\downarrow$} & rel.(\%)~\textcolor{red}{$\downarrow$} \\ \hline
        \multirow{2}{*}{Word/Char} 
        ~ & TextBugger & 0.33 & 44.71 & 34.50 & 0.04 & 38.10 & 10.27 & 0.04 & 34.92 & 6.37 \\ \cline{2-11}
        ~ & HotFlip & 0.33 & 45.90 & 34.60 & 0.05 & 62.69 & 11.91 & 0.04 & 39.68 & 6.27  \\ \cline{2-11}
        ~ & Median & 0.33 & 45.31 & 34.55 & 0.05 & 50.40 & 11.09 & 0.04 & 44.44 & 6.12 \\ \hline
        \multirow{5}{*}{Character} & DeepWordBug & 0.34 & 44.03 & 34.09 & 0.05 & 62.69 & 12.73 & 0.04 & 44.44 & 6.16  \\ \cline{2-11}
        ~ & C1 & 0.57 & 4.61 & 9.14 & 0.13 & 4.55 & 6.37 & 0.07 & -1.59 & 2.57  \\ \cline{2-11}
        ~ & C2 & 0.57 & 4.78 & 11.09 & 0.10 & 22.73 & 9.03 & 0.08 & -17.46 & 4.62  \\ \cline{2-11}
        ~ & C3 & 0.59 & 2.39 & 5.54 & 0.13 & 1.52 & 6.78 & 0.06 & 6.35 & 2.67  \\ \cline{2-11}
        ~ & C4 & 0.60 & 1.02 & 6.78 & 0.11 & 15.15 & 5.75 & 0.07 & -7.94 & 2.36  \\ \cline{2-11}
        ~ & Median & 0.57 & 4.61 & 9.14 & 0.11 & 15.15 & 6.78 & 0.07 & -1.59 & 2.67  \\ \hline
        \multirow{6}{*}{Word}& PWWS & 0.33 & 45.39 & 34.91 & 0.04 & 67.91 & 12.22 & 0.04 & 44.44 & 5.75  \\ \cline{2-11}
        ~ & PSO & 0.33 & 44.88 & 34.60 & 0.04 & 70.90 & 12.63 & 0.04 & 34.92 & 5.95  \\ \cline{2-11}
        ~ & FD & 0.36 & 40.27 & 33.06 & 0.05 & 61.19 & 12.32 & 0.04 & 34.92 & 6.16  \\ \cline{2-11}
        ~ & W1 & 0.58 & 3.92 & 8.52 & 0.11 & 18.18 & 6.16 & 0.07 & -6.35 & 2.46  \\ \cline{2-11}
        ~ & W2 & 0.55 & 7.85 & 10.68 & 0.12 & 12.88 & 5.44 & 0.07 & -11.11 & 3.59  \\ \cline{2-11}
        ~ & W3 & 0.60 & 0.68 & 3.90 & 0.13 & 6.82 & 2.77 & 0.07 & -6.35 & 0.82 \\ \cline{2-11}
        ~ & Median & 0.46 & 24.06 & 21.87 & 0.08 & 39.69 & 9.19 & 0.06 & 14.29 & 4.67  \\ \hline
        \multirow{4}{*}{Sentence} 
        ~& S1 & 0.60 & -0.17 & 2.16 & 0.14 & -4.55 & 1.64 & 0.07 & -1.59 & 1.13 \\ \cline{2-11}
        ~ & S2 & 0.61 & 0.84 & 5.44 & 0.15 & 13.95 & 4.52 & 0.21 & 2.43 & 3.18  \\ \cline{2-11}
        ~ & S3 & 0.60 & 0.17 & 6.26 & 0.13 & 6.06 & 5.54 & 0.07 & -9.52 & 3.08  \\ \cline{2-11}
        ~ & S4 & 0.57 & 4.95 & 9.55 & 0.12 & 14.39 & 7.08 & 0.04 & 41.27 & 4.52  \\ \cline{2-11}
        ~ & Median & 0.60 & 0.51 & 5.85 & 0.14 & 10.01 & 5.03 & 0.07 & 0.42 & 3.13 \\ \hline

    \end{tabular}%
    }
\end{table*}

\begin{table*}[!ht]
    \centering
    \caption{Evaluation of Robustness Metrics on Three Large Language models for Code Using the MBPP Dataset Perturbed in the task description. An upward arrow ($\uparrow$) indicates that a higher value corresponds to better robustness, while a downward arrow ($\downarrow$) indicates the opposite. Codegen-2B-mono shows the best performance in terms of ROBUSTNESS.}
    \label{tab:results_RQ2_4}
     \resizebox{\textwidth}{!}{
     \begin{tabular}{ll|lll|lll|lll}
    \hline
        \textbf{ Perturbations} &~ & \multicolumn{3}{c}{\textbf{codegen-2B-mono}} &\multicolumn{3}{|c|}{\textbf{codegen-2B-multi}} &\multicolumn{3}{c}{\textbf{incoder-1B}}    \\ \hline
        \textbf{Level} & \textbf{Method} & \textbf{passatk}~\textcolor{blue}{$\uparrow$} &\textbf{drop(\%)}~\textcolor{red}{$\downarrow$} & \textbf{rel.(\%)}~\textcolor{red}{$\downarrow$} &\textbf{passatk}~\textcolor{blue}{$\uparrow$} & \textbf{drop(\%)}~\textcolor{red}{$\downarrow$} & \textbf{rel.(\%)}~\textcolor{red}{$\downarrow$} &\textbf{passatk}~\textcolor{blue}{$\uparrow$}& \textbf{drop(\%)}~\textcolor{red}{$\downarrow$} & \textbf{rel.(\%)}~\textcolor{red}{$\downarrow$}  \\ \hline
        \multirow{2}{*}{Word/Char} & TextBugger & 0.32 & 0.00 & 0.00 & 0.09 & 52.64 & 15.40 & 0.05 & 62.36 & 11.29  \\ \cline{2-11}
        ~ & HotFlip & 0.16 & 50.75 & 23.51 & 0.09 & 54.25 & 15.71 & 0.04 & 67.17 & 11.09  \\ \cline{2-11}
        ~ & Median & 0.24 & 25.37 & 11.76 & 0.09 & 53.45 & 15.55 & 0.05 & 64.76 & 11.19  \\ \hline
        \multirow{5}{*}{Character} & DeepWordBug & 0.17 & 45.63 & 22.28 & 0.11 & 42.95 & 14.99 & 0.06 & 55.95 & 11.29  \\ \cline{2-11}
        ~ & C1 & 0.21 & 33.66 & 20.43 & 0.09 & 51.61 & 21.66 & 0.04 & 65.60 & 14.48  \\ \cline{2-11}
        ~ & C2 & 0.19 & 41.10 & 22.07 & 0.09 & 54.30 & 20.74 & 0.05 & 58.40 & 13.24  \\ \cline{2-11}
        ~ & C3 & 0.23 & 26.86 & 16.53 & 0.12 & 39.78 & 15.81 & 0.06 & 54.40 & 12.42  \\ \cline{2-11}
        ~ & C4 & 0.21 & 32.69 & 20.64 & 0.11 & 44.09 & 17.66 & 0.06 & 55.20 & 12.42  \\ \cline{2-11}
        ~ & Median & 0.21 & 33.66 & 20.64 & 0.11 & 44.09 & 18.28 & 0.06 & 55.95 & 12.42  \\ \hline
        \multirow{5}{*}{Word} & PWWS & 0.32 & 0.32 & 0.10 & 0.18 & 4.84 & 10.37 & 0.10 & 19.20 & 5.75  \\ \cline{2-11}
        ~ & PSO & 0.32 & 0.00 & 0.00 & 0.09 & 54.79 & 15.61 & 0.05 & 63.16 & 11.60  \\ \cline{2-11}
        ~ & FD & 0.32 & 0.00 & 0.00 & 0.09 & 52.64 & 15.40 & 0.05 & 62.36 & 11.29  \\ \cline{2-11}
        ~ & W1 & 0.24 & 24.92 & 16.63 & 0.11 & 41.94 & 18.99 & 0.05 & 59.20 & 13.14  \\ \cline{2-11}
        ~ & W2 & 0.19 & 39.16 & 22.79 & 0.10 & 48.39 & 18.17 & 0.03 & 79.20 & 16.22  \\ \cline{2-11}
        ~ & W3 & 0.31 & 0.97 & 3.18 & 0.20 & -3.23 & 3.29 & 0.12 & 4.80 & 2.46 \\ \cline{2-11}
        ~ & Median & 0.32 & 0.32 & 0.10 & 0.10 & 48.39 & 15.61 & 0.05 & 62.36 & 11.60  \\ \hline
        \multirow{5}{*}{Sentence} 
        ~ & S1 & 0.32 & -0.32 & 2.16 & 0.19 & 0.54 & 2.36 & 0.12 & 6.40 & 1.64 \\ \cline{2-11}
        ~ & S3 & 0.40 & 1.52 & 4.27 & 0.27 & 6.25 & 4.27 & 0.18 & 3.23 & 0.61  \\ \cline{2-11}
        ~ & S4 & 0.30 & 4.21 & 6.26 & 0.19 & 2.69 & 6.06 & 0.12 & 7.20 & 4.00  \\ \cline{2-11}
        ~ & S5 & 0.31 & 3.56 & 8.93 & 0.16 & 15.59 & 10.78 & 0.10 & 20.00 & 7.08  \\ \cline{2-11}
        ~ & Median & 0.31 & 2.54 & 5.27 & 0.19 & 4.47 & 5.17 & 0.12 & 6.80 & 2.82 \\ \hline

    \end{tabular}%
    }
\end{table*}

\begin{table}[!ht]
    \centering
    \caption{Wilcoxon signed-ranked test($\alpha$ = 0.05) for RQ2. We applied the Wilcoxon signed-rank test to compare model performance under different perturbation conditions and used Bonferroni correction to adjust for multiple comparisons. The majority of p-values are below 0.05 and 87\% of comparisons show large effect sizes, reinforcing the practical impact of these differences.}
    \label{tab:wilcoxon_RQ2}
    \resizebox{\textwidth}{!}{
    \begin{tabular}{|l|l|l|l|l|l|l|l|l|}
    \hline
        Wilcoxon Statistic & p-value & Bonferroni Correction & Effect Size (r) & ~ & Wilcoxon Statistic & p-value & Bonferroni Correction & Effect Size (r)  \\ \hline
        
        0 & 0.000289619 & 0.004344285 & 0.87988269 & ~ & 0 & 0.000292478 & 0.00438717 & 0.87988269  \\ \hline
        0 & 0.000291332 & 0.00436998 & 0.87988269 & ~ & 0 & 0.000292478 & 0.00438717 & 0.87988269  \\ \hline
        4 & 0.000935091 & 0.014026365 & 0.821223844 & ~ & 0 & 1.53E-05 & 0.0002295 & 0.87988269  \\ \hline
        0 & 0.000291905 & 0.004378575 & 0.87988269 & ~ & 0 & 1.53E-05 & 0.0002295 & 0.87988269  \\ \hline
        0 & 1.53E-05 & 0.0002295 & 0.87988269 & ~ & 0 & 1.53E-05 & 0.0002295 & 0.87988269  \\ \hline
        0 & 0.000291905 & 0.004378575 & 0.87988269 & ~ & 0 & 1.53E-05 & 0.0002295 & 0.87988269  \\ \hline
        46 & 0.148734892 & 2.23102338 & 0.205305961 & ~ & 22 & 0.007904053 & 0.118560795 & 0.557259037  \\ \hline
        0 & 0.00029019 & 0.00435285 & 0.87988269 & ~ & 0 & 1.53E-05 & 0.0002295 & 0.87988269  \\ \hline
        0 & 0.000292478 & 0.00438717 & 0.87988269 & ~ & 0 & 1.53E-05 & 0.0002295 & 0.87988269  \\ \hline
        7 & 0.000993699 & 0.014905485 & 0.77722971 & ~ & 47 & 0.00324568 & 0.0486852 & 0.19064125  \\ \hline
        2 & 0.000416879 & 0.006253185 & 0.850553267 & ~ & 55 & 0.328948975 & 4.934234625 & 0.073323558  \\ \hline
        0 & 0.00029019 & 0.00435285 & 0.87988269 & ~ & 0 & 1.53E-05 & 0.0002295 & 0.87988269  \\ \hline
        67 & 0.677703857 & 10.16555786 & 0.102652981 & ~ & 33 & 0.039535522 & 0.59303283 & 0.395947211  \\ \hline
        0 & 0.000292478 & 0.00438717 & 0.87988269 & ~ & 0 & 1.53E-05 & 0.0002295 & 0.87988269  \\ \hline
        0 & 0.000292478 & 0.00438717 & 0.87988269 & ~ & 0 & 1.53E-05 & 0.0002295 & 0.87988269 \\ \hline
    \end{tabular}
    }
\end{table}

\noindent\textbf{Results and Analysis:} Tables~\ref{tab:results_RQ2_1} and~\ref{tab:results_RQ2_2} show the results of our experiments with models on the HumanEval dataset, subjected to the three levels of perturbations mentioned earlier. Similarly, Tables~\ref{tab:results_RQ2_3} and~\ref{tab:results_RQ2_4} display the results of experiments with models on the MBPP dataset under the same perturbation levels.

The analysis of the results obtained from both datasets for Large Language models for Code offers valuable insights into model robustness across different perturbation levels. For all six models,
the robust pass (RP) does not vary significantly in HumanEval datasets (e.g. the RP values of codegen-2B-mono model under sentense-level, word-level and character-level perturbations are 0.25, 0.23, 0.23.) due to the inherent simplicity or standardized nature of the tasks within this dataset, which may be less susceptible to certain types of perturbations. 
Therefore, we focus on comparing RDs in detail. The general observation is that most models (e.g. the RD values of codegen-2B-mono model under sentense-level, word-level and character-level perturbations are 4.88\%, 24.08\%, 12.20\%) exhibit the highest robustness to sentence-level perturbations and the lowest to word-level perturbations. 
However, for MBPP datasets, some models show best robustness under the sentence and second under character levels (e.g. RP values of Meta-Llama-3-8B-Instruct are 0.60, 0.57, RD values are 0.51\%, 4.61\%), with slightly better performance at the sentence level. The varied responses to different perturbations in task descriptions can be attributed to factors such as the type of the perturbations, the training data and techniques, the model architecture, context handling, error correction mechanisms, and sensitivities to text structure and formatting. Understanding these factors is crucial for developing more robust Large Language models for Code that can effectively handle a wide range of perturbations in task descriptions. For multi-level perturbations (Word-Char level), the performance tends to degrade similarly to that observed at the word level.

For the codegen family, ``mono'' models consistently demonstrate superior robustness to every perturbation compared to ``multi'' models. This advantage is due to their specialized focus, reduced complexity, efficient resource utilization, access to rich and consistent training data, contextual consistency, and optimized architectures. These factors collectively enable to handle various perturbations more effectively, resulting in superior performance and robustness.

\newpage
\begin{answer*}{ (RQ2)}{}
Perturbations at varying linguistic levels—word, character, and sentence—in task descriptions significantly impact the functional correctness of the code generated by models. Word-level modifications, in particular, severely impair most models' task comprehension, while sentence-level perturbations show comparatively greater robustness. These findings suggest that both the type of perturbation (word, character, or sentence) and the intensity of perturbation critically influence a model's ability to generate functionally correct code. Also we find that mono language models often performance better than multi language models.
\end{answer*}

\noindent \textbf{\rqthree}

\noindent\minor{\textbf{Scope:} Perturbations affect inline or block comments embedded within code that are visible to the model during inference.}

\noindent\textbf{Objective:}
This research question seeks to evaluate the impact of perturbations introduced at different levels—character, word, and sentence—within the comments of code on the functional correctness of output generated by Large Language models for Code. The goal is to determine how modifications to comments, which typically provide context or explain code logic without influencing its execution, affect the model's ability to generate functionally correct and semantically accurate code. The findings will offer insights into the robustness of Large Language models for Code to alterations in non-executable parts of the code and underscore the importance of comments in the automated code generation process.

\noindent\textbf{Approach:} \edit{Our dataset does not contain pre-existing comments, making direct modifications to comments unfeasible. Instead, we introduce a controlled perturbation by injecting comments into code, allowing us to examine how LLM4Code respond to the presence of comments where none previously existed. This perturbation reflects real-world scenarios where developers add comments to improve readability or provide additional context. Although this differs from modifying existing comments, it still serves as a meaningful investigation into the impact of comments on LLM4Code robustness.}
\edit{To evaluate RQ3, we utilized the Doc2Comments perturbation method from ReCode, which converts structured docstrings into inline or block comments while preserving content. This transformation enabled us to create a test dataset where task descriptions appeared as comments instead of docstrings, allowing us to analyze the impact of documentation format variations on LLM4Code performance.} We utilized the doc2comments perturbation method from ReCode to create an additional test dataset. The definition of these perturbations are shown in Table~\ref{tab:reclassification_for_RQ3}. As shown in Listing~\ref{lst:doc2com}, the description from lines 3 to 4 was converted into comments by adding the symbol \# at the beginning. 
Specifically, a single-line comment was inserted at a randomly chosen location within each function body (as illustrated in Listing~\ref{lst:randominsert}, where ``\#This is a randomly inserted comment'' was added in line 11). Given the concise nature of the functions in these datasets, only one comment was added per function, and the relevance of the comment content to the function was not considered in this process. Following the introduction of these perturbations, the models were evaluated using the modified datasets, allowing us to assess the robustness of the models based on the three previously defined metrics. \edit{due to a limited sample size in RQ3, statistical significance could not be established, and effect size calculations were omitted to avoid misleading conclusions}


\noindent\begin{minipage}{.45\textwidth}
\begin{lstlisting}[escapechar=!, language=Python, label=lst:doc2com, caption=``doc2comment'' perturbations. The description from lines 3-4 is commented on by adding the symbol \# at the beginning]
def rolling_max(numbers):
    """
    !\colorbox{yellow}{\#}!From a given list of integers, generate a list of rolling maximum element found until given moment in the sequence.
    >>> rolling_max([1, 2, 3, 2, 3, 4, 2])
    [1, 2, 3, 3, 3, 4, 4]
    """
    running_max = None
    result = []

    for n in numbers:
        if running_max is None:
            running_max = n
        else:
            running_max = max(running_max, n)
        result.append(running_max)
    return result
\end{lstlisting}
\end{minipage}\hfill
\begin{minipage}{.5\textwidth}
\begin{lstlisting}[escapechar=!, language=Python, label=lst:randominsert, caption=``randominsertcomments'' perturbations. Showing inserted comments after a code at line 11: ``\#This is a randomly inserted in line'']
def rolling_max(numbers):
    """
    From a given list of integers, generate a list of rolling maximum element found until given moment in the sequence.
    >>> rolling_max([1, 2, 3, 2, 3, 4, 2])
    [1, 2, 3, 3, 3, 4, 4]
    """
    running_max = None
    result = []
    for n in numbers:
        if running_max is None:
            running_max = n !\colorbox{yellow}{\#This is a randomly inserted comment}!
    return result
        else:
            running_max = max(running_max, n)
        result.append(running_max) 
\end{lstlisting}
\end{minipage}\hfill

\begin{table*}[!ht]
\centering
\caption{Definition of comment perturbations}
\label{tab:reclassification_for_RQ3}
\resizebox{\linewidth}{!}{
\begin{threeparttable}
    \begin{tabular}{l|l|l} 
\hline
\hline
\textbf{Perturbation Method}          & \textbf{Definition}   & \textbf{Classification }\\
\hline
\hline
doc2comment &converts structured docstrings into inline comments &Sentence-Level\\
\hline
randominsertcomments &randomly insert unrelated comments &Sentence-Level\\
\hline
\end{tabular}
\end{threeparttable}
}
\end{table*}

\noindent\textbf{Results and Analysis:} Tables~\ref{tab:results_RQ3_1}, ~\ref{tab:results_RQ3_2}, ~\ref{tab:results_RQ3_3}, and ~\ref{tab:results_RQ3_4} demonstrate that 
\edit{Models exhibit notably lower robustness to comment perturbations compared to code and task description perturbations. This suggests that LLM4Code rely heavily on contextual information provided by comments, making them more susceptible to adversarial modifications in documentation.}
This decrease in robustness could be explained by the fact that comments do not follow the same syntactic or semantic rules and can create ambiguity or disrupt the model's comprehension of the code structure. Randomly inserted comments might obscure critical code sections or introduce noise that confuses the model's ability to maintain logical code flow.

Regarding the difference between doc2comment and random comment insertions, the impact of random comments is generally more severe. Doc2comments are more structured and relevant to the code, hence having a relatively smaller impact. Random comments, on the other hand, lack context and relevance, which can disrupt the model's ability to focus on essential parts of the code, leading to greater performance degradation. This difference highlights the importance of comment relevance and context in determining the level of disruption caused by comment perturbations.

This finding suggests that both the location and content of comment insertions could significantly impact the quality of the generated code. Further analysis revealed that models within the ``mono'' architecture category generally outperformed those in the ``multi'' category when handling comment perturbations. Mono language models are typically designed for a single task or language, making them more specialized and likely better at maintaining robustness under simpler perturbations like comments. In contrast, multi language model architectures may have broader applications but struggle with maintaining robustness in specific contexts like comment perturbations, which introduces additional complexity.

However, within the mono category, performance varied across models. For example, the robustness to comment perturbations was even more compromised when applied to the HumanEval dataset (RP values of gpt-j-6B are 0.21, 0.02 in HumanEval, and 0.17, 0.11 in MBPP), indicating a broader challenge across datasets. This underlines the necessity for models to not only process code accurately but also to handle non-code elements like comments effectively. The impact of comment perturbations shows that integrating better noise-handling or context-aware mechanisms could be beneficial in enhancing model robustness. Furthermore, improving model training to distinguish between critical code and auxiliary information (such as comments) could lead to more resilient Large Language models for Code across varying datasets.



\begin{table*}[!ht]
    \centering
  \caption{Evaluation of Robustness Metrics on Three Code-Task Models Using the Humaneval Dataset Perturbed with Comments. Perturbations include \texttt{doc2comments} (doc2Coms) and \texttt{Randominsertcomments} (RandComs). An upward arrow ($\uparrow$) indicates that a higher value corresponds to better robustness, while a downward arrow ($\downarrow$) indicates the opposite.}
    \label{tab:results_RQ3_1}
     \resizebox{\textwidth}{!}{
     \begin{tabular}{ll|lll|lll|lll}
    \hline
        Perturbations& ~ & \multicolumn{3}{c}{Meta-Llama-3-8B-Instruct} &  \multicolumn{3}{|c|}{gpt-j-6b} & \multicolumn{3}{c}{CodeLlama-7b-hf}   \\ \hline
        
        \textbf{Level} & \textbf{Method} & \textbf{passatk}~\textcolor{blue}{$\uparrow$} &\textbf{drop(\%)}~\textcolor{red}{$\downarrow$} & \textbf{rel.(\%)}~\textcolor{red}{$\downarrow$} &\textbf{passatk}~\textcolor{blue}{$\uparrow$} & \textbf{drop(\%)}~\textcolor{red}{$\downarrow$} & \textbf{rel.(\%)}~\textcolor{red}{$\downarrow$} &\textbf{passatk}~\textcolor{blue}{$\uparrow$}& \textbf{drop(\%)}~\textcolor{red}{$\downarrow$} & \textbf{rel.(\%)}~\textcolor{red}{$\downarrow$} \\ \hline
        
        \multirow{2}{*}{Sentence} & 
        doc2Coms & 0.63 & 5.50 & 12.20 & 0.21 & -9.68 & 7.93  & 0.20 & 21.95 & 10.37 \\ \cline{2-11}
        
        ~ & RandComs & 0.58 & 0.00 & 4.88  & 0.02 & 55.56 & 4.27 & 0.10 & 10.53 & 1.22\\ \hline

    \end{tabular}%
    }
\end{table*}

\begin{table*}[!ht]
    \centering
    \caption{Evaluation of Robustness Metrics on Three Large Language models for Code Using the Humaneval Dataset Perturbed with Comments. Perturbations include \texttt{doc2comments} (doc2Coms) and \texttt{Randominsertcomments} (RandComs). An upward arrow ($\uparrow$) indicates that a higher value corresponds to better robustness, while a downward arrow ($\downarrow$) indicates the opposite.}
    \label{tab:results_RQ3_2}
     \resizebox{\textwidth}{!}{
     \begin{tabular}{ll|lll|lll|lll}
    \hline
        \textbf{ Perturbations} &~ & \multicolumn{3}{c}{\textbf{codegen-2B-mono}} &\multicolumn{3}{|c|}{\textbf{codegen-2B-multi}} &\multicolumn{3}{c}{\textbf{incoder-1B}}   \\ \hline
        
         \textbf{Level} & \textbf{Method} & \textbf{passatk}~\textcolor{blue}{$\uparrow$} &\textbf{drop(\%)}~\textcolor{red}{$\downarrow$} & \textbf{rel.(\%)}~\textcolor{red}{$\downarrow$} &\textbf{passatk}~\textcolor{blue}{$\uparrow$} & \textbf{drop(\%)}~\textcolor{red}{$\downarrow$} & \textbf{rel.(\%)}~\textcolor{red}{$\downarrow$} &\textbf{passatk}~\textcolor{blue}{$\uparrow$}& \textbf{drop(\%)}~\textcolor{red}{$\downarrow$} & \textbf{rel.(\%)}~\textcolor{red}{$\downarrow$} \\ \hline
        
        \multirow{2}{*}{Sentence} & doc2Coms & 0.34 & 16.42 & 6.71 & 0.31 & -4.08 & 6.10 &  0.15 & 22.58 & 5.49  \\ \cline{2-11}
        
        ~ & RandComs & 0.24 & 40.30 & 18.90 & 0.12 & 59.18 & 21.34 & 0.09 & 51.61 & 13.41 \\ \hline

    \end{tabular}%
    }
\end{table*}

\begin{table*}[!ht]
    \centering
  \caption{Evaluation of Robustness Metrics on Three Code-Task Models Using the MBPP Dataset Perturbed with Comments. Perturbations include \texttt{doc2comments} (doc2Coms) and \texttt{Randominsertcomments} (RandComs). An upward arrow ($\uparrow$) indicates that a higher value corresponds to better robustness, while a downward arrow ($\downarrow$) indicates the opposite.}
    \label{tab:results_RQ3_3}
     \resizebox{\textwidth}{!}{
     \begin{tabular}{ll|lll|lll|lll}
    \hline
        Perturbations& ~ & \multicolumn{3}{c}{Meta-Llama-3-8B-Instruct} &  \multicolumn{3}{|c|}{gpt-j-6b} & \multicolumn{3}{c}{CodeLlama-7b-hf}   \\ \hline
        
        \textbf{Level} & \textbf{Method} & \textbf{passatk}~\textcolor{blue}{$\uparrow$} &\textbf{drop(\%)}~\textcolor{red}{$\downarrow$} & \textbf{rel.(\%)}~\textcolor{red}{$\downarrow$} &\textbf{passatk}~\textcolor{blue}{$\uparrow$} & \textbf{drop(\%)}~\textcolor{red}{$\downarrow$} & \textbf{rel.(\%)}~\textcolor{red}{$\downarrow$} &\textbf{passatk}~\textcolor{blue}{$\uparrow$}& \textbf{drop(\%)}~\textcolor{red}{$\downarrow$} & \textbf{rel.(\%)}~\textcolor{red}{$\downarrow$} \\ \hline
        
        \multirow{2}{*}{Sentence} & 
        doc2Coms & 0.63 & -2.18 & 7.49 & 0.17 & 2.33 & 6.78 & 0.25 & -20.39 & 9.65 \\ \cline{2-11}
        
        ~ & RandComs & 0.58 & 3.75 & 5.54 & 0.11 & 23.13 & 7.91 & 0.07 & -15.87 & 5.13 \\ \hline

    \end{tabular}%
    }
\end{table*}

\begin{table*}[!ht]
    \centering
    \caption{Evaluation of Robustness Metrics on Three Large Language models for Code Using the MBPP Dataset Perturbed with Comments. Perturbations include \texttt{doc2comments} (doc2Coms) and \texttt{Randominsertcomments} (RandComs). An upward arrow ($\uparrow$) indicates that a higher value corresponds to better robustness, while a downward arrow ($\downarrow$) indicates the opposite.}
    \label{tab:results_RQ3_4}
     \resizebox{\textwidth}{!}{
     \begin{tabular}{ll|lll|lll|lll}
    \hline
        \textbf{ Perturbations} &~ & \multicolumn{3}{c}{\textbf{codegen-2B-mono}} &\multicolumn{3}{|c|}{\textbf{codegen-2B-multi}} &\multicolumn{3}{c}{\textbf{incoder-1B}}   \\ \hline
        
         \textbf{Level} & \textbf{Method} & \textbf{passatk}~\textcolor{blue}{$\uparrow$} &\textbf{drop(\%)}~\textcolor{red}{$\downarrow$} & \textbf{rel.(\%)}~\textcolor{red}{$\downarrow$} &\textbf{passatk}~\textcolor{blue}{$\uparrow$} & \textbf{drop(\%)}~\textcolor{red}{$\downarrow$} & \textbf{rel.(\%)}~\textcolor{red}{$\downarrow$} &\textbf{passatk}~\textcolor{blue}{$\uparrow$}& \textbf{drop(\%)}~\textcolor{red}{$\downarrow$} & \textbf{rel.(\%)}~\textcolor{red}{$\downarrow$} \\ \hline
        
        \multirow{2}{*}{Sentence} & doc2Coms & 0.34 & 16.67 & 11.59 & 0.29 & 2.08 & 5.49 & 0.15 & 22.58 & 4.27  \\ \cline{2-11}
        
        ~ & RandComs & 0.32 & 0.00 & 0.00 & 0.17 & 11.29 & 5.65 & 0.11 & 16.80 & 4.62 \\ \hline

    \end{tabular}%
    }
\end{table*}


\newpage
\begin{answer*}{ (RQ3)}{}
Perturbations introduced through comments in code can significantly impact the functional correctness of the code generated by models, with models exhibiting varied levels of sensitivity, especially to randomly inserted comments. This suggests that both the placement and content of comments can adversely affect the model's interpretation and output quality, with some models demonstrating better robustness than others. Also we find that mono language models often performance better than multi language models.
\end{answer*}

\section{Discussion } \label{sec:discussion}

\minor{
Our study provides a rigorous and multifaceted examination of the robustness of Large Language Models for Code (LLM4Code) under structured adversarial perturbations. Consistent with prior research, notably ReCode~\cite{wang2022recode}, our results confirm that LLM4Code models exhibit varying degrees of vulnerability depending on the granularity and surface of the perturbation. Furthermore, our extended evaluation across a broader perturbation taxonomy and model family reveals additional, significant insights. For example, under the BackTranslation perturbation for the codegen-2B-mono model, we recorded robustness values of RP@1 = 0.25, RD@1 = 0.00\%, and RR@1 = 1.22\%, in contrast to ReCode's reported 0.213, 7.89\%, and 4.27\%, respectively. These divergences reflect the sensitivity of LLMs to experimental variables such as dataset versions, tokenizers, or decoding strategies, underscoring the pressing need for standardized evaluation frameworks when comparing robustness studies.\\
Our comparative analysis shows that sentence-level perturbations are generally met with higher resilience, whereas word-level perturbations—particularly involving synonym substitution, variable renaming, or misleading entity replacement—consistently induce substantial performance degradation across models. This highlights a core vulnerability in LLM4Code: many models exhibit an over-reliance on surface-level token patterns rather than semantic abstractions, limiting their capacity to generalize across minor lexical variations. Such findings are aligned with CWEval~\cite{peng2025cweval} and YABLoCo~\cite{valeev2025yabloco}, which report similar degradation in the face of subtle prompt alterations, especially when critical tokens or semantic markers are obfuscated.\\
From a statistical standpoint, the Wilcoxon signed-rank test conducted in RQ1 and RQ2 indicates that a majority of p-values fall below 0.05, while over 87\% of comparisons reveal large effect sizes. This strongly supports the robustness of our empirical observations and validates the practical impact of adversarial perturbations on model output fidelity. These results reinforce the conclusion that robustness is not evenly distributed across perturbation types or model architectures, and that fine-grained textual changes, although seemingly innocuous, can lead to meaningful functional regressions.\\
A noteworthy trend observed in our study is the consistent outperformance of monolingual models over multilingual ones in robustness scenarios. Monolingual models such as codegen-2B-mono demonstrated superior resistance to adversarial modifications due to their specialized training on language-specific corpora (e.g., Python). In contrast, multilingual models, tasked with balancing performance across diverse programming languages, showed lower robustness due to their generalized tokenization and broader training objectives. This echoes findings in multilingual NLP literature~\cite{raffel2020exploring,lin2022random} and robustness studies in educational contexts~\cite{s2024investigating}, emphasizing the trade-off between generality and specialization in model training.\\
Importantly, our structured use of ReCode and OpenAttack enables systematic stress-testing of both code and natural language input surfaces. This dual evaluation is vital, as many vulnerabilities lie not only in the code snippet itself but also in peripheral elements such as task descriptions and inline comments—components frequently encountered in real-world development prompts. Our taxonomy-driven perturbation framework reveals that even minor changes in comments or docstrings can steer model generation toward unintended or incorrect outputs, thereby representing both functional and security risks.\\
From a broader perspective, our findings carry significant practical implications for the secure deployment of LLM4Code in software development. In high-assurance domains such as healthcare, finance, or safety-critical systems, the brittleness of LLMs to minor prompt variations can introduce subtle but catastrophic failures. The robustness metrics introduced in our study—RP@k, RD@k, and RR@k—enable practitioners to quantify and benchmark model stability under realistic adversarial scenarios, thus forming the basis for robust tooling, verification protocols, and fine-tuning strategies.\\
The results also reveal promising directions for improving LLM4Code robustness. Techniques such as adversarial data augmentation, perturbation-aware fine-tuning, and reinforcement learning from human feedback offer scalable pathways to enhance fault tolerance. Moreover, the integration of pre-processing filters and ensemble verification frameworks can provide defense-in-depth by detecting and mitigating adversarial inputs before execution. These strategies align with trends in adversarial defense across broader machine learning domains and offer tangible next steps for model developers.\\
While our perturbation suite is extensive, it is not exhaustive. Future work should extend our taxonomy to cover adversarial attack types such as prompt injection, long-context inconsistency, or logic obfuscation—areas currently underexplored in LLM4Code research. Likewise, developing hybrid evaluation pipelines that incorporate both static and dynamic analysis of generated code can deepen our understanding of model robustness, especially in edge cases where functional correctness and semantic alignment diverge.\\
In addition, our study not only demonstrates the fragility of current LLM4Code systems under adversarial conditions but also provides the methodological foundation and empirical insights needed to build more resilient AI-assisted programming tools. The findings highlight the critical need for robustness-aware evaluation in LLM development and establish a roadmap for both academic inquiry and industrial adoption in securing the next generation of intelligent coding systems.
}

\section{Implications } \label{sec:implication}

\minor{
Our findings provide important implications for multiple stakeholders involved in the development, deployment, and oversight of LLM4Code. Drawing from empirical observations and robustness evaluation, we now outline how this study informs best practices, risk management, and future research priorities.\\
\textbf{For Software Developers and Practitioners:} The study highlights the non-trivial impact that minor input changes, such as a paraphrased docstring or a typo in a comment, can have on model output. In practice, this suggests that LLM-based code generation tools must be integrated with caution, particularly in production or mission-critical environments. Developers should not assume that LLMs will behave reliably under benign variations in prompts. Instead, robust prompt engineering guidelines and code review procedures should be mandated, especially when LLM outputs are directly used in development workflows.\\
\textbf{For Software Testers and QA Engineers:} Our perturbation classification introduces a systematic way to test robustness across character, word, and sentence levels, and across different prompt surfaces (code, comments, and descriptions). Sentence-level perturbations were found to cause minimal degradation, while word-level changes significantly reduced functional correctness. These findings advocate for the integration of adversarial prompt testing into automated testing pipelines, enabling early detection of brittleness in LLM-supported development tools.\\
\textbf{For Ethical Hackers and Security Analysts:} The demonstrated vulnerability of LLMs to word-level and character-level perturbations in natural language inputs indicates tangible risk surfaces for adversarial exploitation. Our structured perturbation types mirror plausible real-world attack vectors such as logic manipulation, backdoor injection, or obfuscation, thereby operationalizing the threat model discussed in Section~\ref{sec:threat-model}. These findings inform red-teaming strategies, helping ethical hackers design targeted probes to evaluate model robustness and system resilience under adversarial scenarios.\\
\textbf{For LLM4Code Developers and AI Scientists:} From an architectural and training standpoint, the evidence suggests that current models struggle with generalizing across semantically equivalent but lexically varied prompts. This over-reliance on surface-level token patterns instead of deeper semantic comprehension suggests that robustness training must go beyond data scale. Fine-tuning models on adversarially perturbed inputs or explicitly optimizing for robustness metrics such as RP@k and RR@k could foster more semantically aligned generation. Additionally, our performance comparisons between mono and multi-language models imply that training specialization enhances robustness, offering a clear direction for future model development.\\
\textbf{For Researchers and Tool Builders:} The multi-dimensional evaluation framework introduced in this study—including perturbation taxonomy and robustness metrics—serves as a practical toolkit for benchmarking LLM4Code under realistic adversarial settings. Our reproducible evaluation protocols, derived from systematic perturbations using tools like ReCode and OpenAttack, offer a replicable basis for comparative research. These can support further studies aimed at adversarial defenses, model interpretability, and safe deployment mechanisms.\\
\textbf{For Software Engineering Managers and Decision-Makers:} The variation in model performance across perturbation types and input surfaces reinforces that not all LLMs offer the same level of resilience. Managers must prioritize robustness testing and risk profiling during vendor selection or toolchain integration. This includes evaluating not only the raw performance of LLMs but also their adversarial stability as a key criterion for tool adoption in regulated or high-assurance environments.\\
In particular, this study confirms that perturbation type, prompt surface, and model architecture are variables that significantly influence LLM robustness. By aligning our experimental design with the threat model and diverse input perturbations, we offer practical insights for secure, responsible, and performance-aware use of LLM4Code across software engineering domains. These implications emphasize that robustness is not merely a research concern but a cross-functional responsibility shared across the software lifecycle, from design and testing to deployment and governance.
}

\section{Limitations and Future Work }\label{sec:limi}
\minor{
Despite providing a comprehensive robustness evaluation of LLM4Code models under structured adversarial perturbations, our study still presents limitations and valuable opportunities for future research.
\subsection{Limitations}
First, the perturbation space explored in this work is constrained by the capabilities of ReCode and OpenAttack, which primarily support character-, word-, and sentence-level perturbations across code, docstrings, and comments. While this granularity covers common real-world errors and adversarial strategies, it does not encompass deeper semantic threats such as logic refactoring, inter-procedural modifications, or multi-module code interactions. These limitations restrict the generalizability of the findings to more complex software systems. 
Second, our evaluation includes only publicly accessible LLM4Code models with fewer than 8 billion parameters. Although this design choice facilitates reproducibility, it excludes leading commercial models (e.g., GPT-4, Claude), limiting our ability to assess robustness at the frontier of LLM capabilities. 
Third, we evaluate robustness primarily through functional correctness, as measured by test case pass rates. While effective for detecting execution-level failures, this method may overlook subtler defects related to security vulnerabilities, code quality, and developer intent. More holistic assessments incorporating maintainability, readability, or security metrics were beyond the scope of this study. 
Fourth, all experiments were conducted on Python-based datasets. As a result, the insights derived may not fully translate to other programming languages, which differ in syntax, semantics, and idiomatic usage. This language-specific focus restricts the breadth of the conclusions. 
Fifth, the training and internal configurations of evaluated models remain opaque, limiting our ability to analyze robustness from an architectural or training-data perspective. This black-box limitation hinders deeper attribution of observed vulnerabilities to model design choices. 
%
%
Last, our results underscore the need for formal robustness benchmarks and threat models tailored to LLM4Code. 
\subsection{Future Work} \label{sec:future-work}
Future work should develop robustness-aware training pipelines to mitigate model brittleness against word-level and semantic perturbations. Three synergistic directions are promising: (1) adversarial data augmentation with taxonomy-defined perturbations to systematically expand training distributions; (2) contrastive learning to enforce invariance across semantically equivalent inputs; and (3) curriculum-based fine-tuning that progressively increases adversarial stress while incorporating perturbation-informed consistency losses—explicitly penalizing output deviations under attack. This unified approach would reduce reliance on superficial token-level cues while jointly optimizing for task accuracy and perturbation robustness. 
Additionally, the development of hybrid evaluation pipelines that assess functional correctness, semantic fidelity, and code quality in tandem will offer a more holistic perspective on model robustness. Future studies could integrate static analysis (e.g., code smells, maintainability metrics), dynamic testing (e.g., mutation testing, fuzzing), and human-in-the-loop review to expose multi-dimensional vulnerabilities and edge cases not captured by simple pass/fail testing alone.
To scale this assessment across real-world scenarios, researchers should develop standardized robustness leaderboards and certification metrics. These benchmarks would help track model progress under diverse adversarial conditions and foster reproducibility across studies.\\
Moreover, evaluation frameworks should consider the roles and needs of different user personas, ranging from novice developers relying heavily on AI assistance to expert programmers and security testers, ensuring that robustness insights translate to usable guidance across contexts.\\
Future work should also position robustness as a first-class concern throughout the LLM4Code lifecycle. This includes designing threat models aligned with software engineering risks, defining robustness-by-design architectures, and building real-time monitoring and alerting systems that detect fragility in deployed code generation environments. The creation of adversarially resilient ecosystems—such as robust code review platforms, adaptive IDEs, and automated patching workflows—remains an open but essential frontier.\\
Ultimately, robustness cannot remain an afterthought. It must be embedded into training protocols, deployment pipelines, and governance structures if LLM4Code systems are to become not only performant but also secure, interpretable, and trustworthy tools in modern software engineering.
}

\section{Threats to Validity} \label{sec:threats}
Throughout this empirical study, we followed the guidelines of Wohlin et al.~\cite{wohlin2012experimentation} and Juristo \& Moreno~\cite{juristo2013basics} to rigorously address and mitigate various threats to validity. We place particular emphasis on controlling for construct, internal, external, and conclusion validity, as well as ensuring the reliability of our results.

\textbf{Construct Validity:} addresses the suitability of the metrics and methodologies employed in evaluating Large Language models for Code. The selected metrics may not fully encapsulate the complexity of code generation tasks or the nuances of certain adversarial attacks. For instance, automated testing frameworks might miss deeper semantic errors that only arise during execution or in specific operational contexts. To ensure construct validity, continuous refinement of evaluation tools is essential to capture the multifaceted nature of code generation. Also, concerning the results provided by ReCode, we observe that discrepancies may arise from variations in experimental settings, underscoring the need for consistency in robustness testing environments to guarantee the comparability of research outcomes.

\textbf{Internal Validity:} refers to the extent to which our study accurately establishes a causal relationship between adversarial attacks and model performance. One threat is that the perturbation techniques used might not cover all possible adversarial scenarios. Additionally, the parameters for generating adversarial inputs, such as the degree of perturbation, could influence model performance and introduce bias. To mitigate these threats, we employed a systematic approach to selecting high-success-rate perturbation methods from established toolkits. Variations in performance metrics between our study and the ReCode project highlight the sensitivity of outcomes to experimental conditions, underscoring the need for meticulous control over these conditions to ensure internal validity.

\textbf{External Validity:} pertains to the generalizability of our findings. A primary threat is the use of specific datasets, which may not represent the full range of real-world software development challenges. While HumanEval and MBPP are reputable benchmarks, they might not capture all coding patterns and nuances present in the industry. Future work could address this by extending evaluations to include more diverse datasets and real-world coding problems.

\textbf{Conclusion Validity:} concerns the degree to which the conclusions drawn from our study are reasonable. This can be threatened by low statistical power or the risk of overfitting models to the datasets and perturbation methods used. Ensuring reproducibility through cross-validation, robust statistical analysis, and sharing of code and data can help mitigate these risks.

\textbf{Reliability:} 
Another consideration is the reliability of the adversarial attack generation process. If the process for creating perturbations is inconsistent or includes uncontrolled random elements, it could lead to variability in the quality of the adversarial examples generated, affecting the reliability of the model robustness evaluation.

\section{Conclusions} \label{sec:conclusions}

Our research provides a comprehensive analysis of model robustness against different levels of perturbations—character, word, sentence, and within comments—across various models and datasets. We found that models exhibit varying sensitivities to perturbations, influenced by both the nature of the perturbation and the model's architecture. Models maintain higher functional correctness against sentence-level perturbations. 
while word-level perturbations reveal the most vulnerabilities. This trend is consistent across different model sizes and configurations, highlighting the importance of model architecture in determining resilience to perturbations.

The introduction of comments, particularly random insertions, significantly impacts model performance, underscoring the nuanced role non-executable code components play in model interpretation. The differential performance between mono and multi-language models suggests that model design decisions profoundly influence performance of robustness, with mono-language models often outperforming multi-language models.
Hence, advocating for a tailored approach in applications demanding high robustness. The variability in model performance across different datasets highlights the substantial challenge of achieving cross-dataset robustness in Large Language models for Code. Models that perform well on one dataset may falter on another, suggesting that the specific characteristics of each dataset significantly influence model performance. Addressing this challenge requires adaptive models capable of adjusting to the unique features of each dataset.

Future research will focus on enhancing the robustness of Large Language models for Code against a variety of perturbations by exploring advanced pre-processing strategies and fine-tuning techniques. We will also evaluate these approaches across a broader range of programming languages and datasets to ensure the resilience of the model. By developing and integrating new metrics that more accurately reflect performance under perturbation, we aim to advance the robustness of Large Language models for Code, improving their reliability and utility in real-world applications.

\begin{acknowledgements}
This work was supported by: Fonds de Recherche du Québec (FRQ), the Canadian Institute for Advanced Research (CIFAR) as well as the DEEL project CRDPJ 537462-18 funded by the Natural Sciences and Engineering Research Council of Canada (NSERC) and the Consortium for Research and Innovation in Aerospace in Québec (CRIAQ), together with its industrial partners Thales Canada inc, Bell Textron Canada Limited, CAE inc and Bombardier inc.
\end{acknowledgements}

\section*{Declaration} \label{sec:Declaration}
\noindent\textbf{Ethical approval} This study did not involve any human participants or animals and therefore did not require ethical approval.

\noindent\textbf{Informed consent} Written informed consent was obtained from all individual participants included in the study.

\noindent\textbf{Author Contributions} Yang Liu conceptualized and designed the study. Yang Liu and Armstrong contributed to writing the draft. Armstrong provides mentorship and formal analysis. All the authors contributed to the manual verification and drafted the manuscript. All authors contributed to reviewing and approving the final manuscript.

\noindent\textbf{Data Availability Statement} The replication package, including our datasets, is available publicly at: 
\url{https://github.com/hugh58623/Robustness-testing}

\noindent\textbf{Conflict of Interests/Competing Interests} The authors declared that they have no conflict of interest.

\noindent\textbf{Clinical Trial Number:} not applicable.

\bibliographystyle{spphys}

\bibliography{references}





\section*{Author Biography}
\begin{wrapfigure}[6]{l}{0.2\textwidth}
    \vspace{-20pt}
    \includegraphics[width=1.5in,height=1.9in,clip,keepaspectratio]{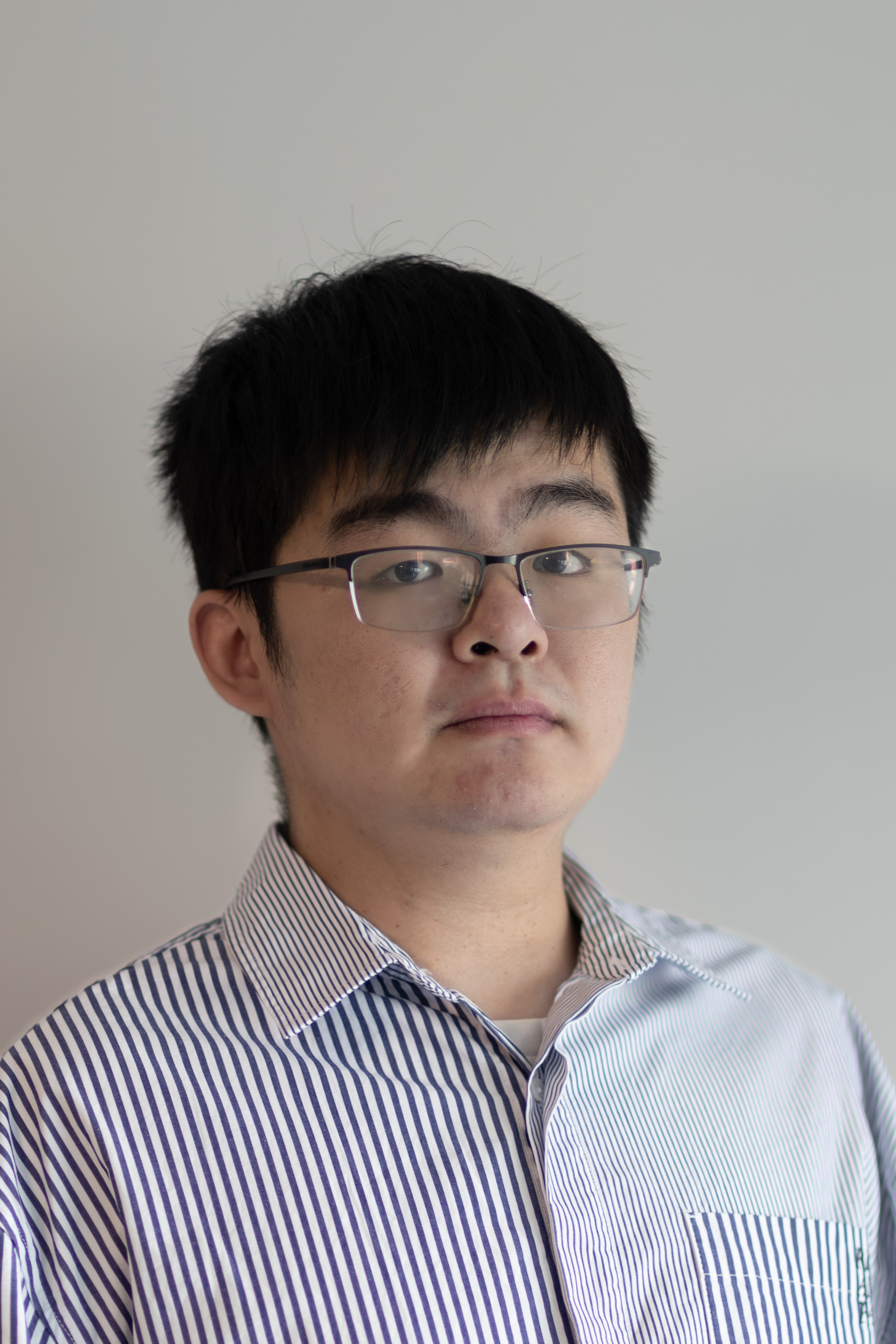}
    \vspace{-10pt}
\end{wrapfigure}
\textbf{Yang Liu} is a Ph.D. candidate in the Department of Computer and Software Engineering at Polytechnique Montréal. He received his M.Sc. degree from The George Washington University (USA) and his B.Eng. degree from the University of Electronic Science and Technology of China. His current research focuses on evaluating and enhancing the robustness of large language models for code generation, with particular emphasis on software security and reliability.

\vspace{40pt}
\begin{wrapfigure}[10]{l}{0.23\textwidth}
    \includegraphics[width=1.5in,height=1.9in,clip,keepaspectratio]{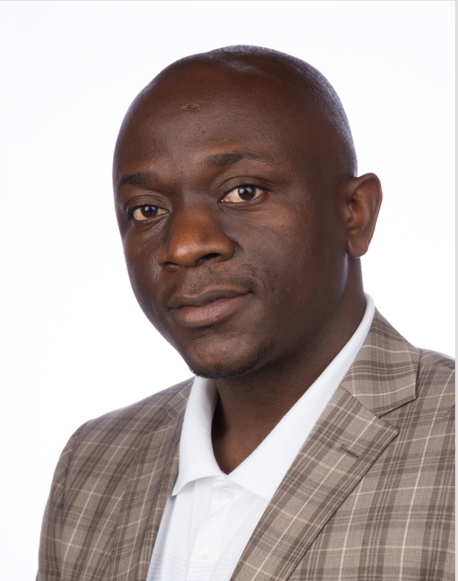}
    \vspace{-10pt}
\end{wrapfigure}
\textbf{Armstrong Foundjem}, P.h.D. is a multidisciplinary research scientist advancing AI/software ecosystem sustainability, the trustworthiness of AI safety-critical systems, cybersecurity, and affective computing. His work bridges software engineering with state-of-the-art machine learning applications—including foundational models, AIWare, and Agentware—to enable scalable, responsible innovation. Foundjem applies socio-technical data science techniques on large-scale ecosystem repositories to extract actionable insights that inform policy and guide decision-making in complex software ecosystems. A passionate advocate for open-source solutions, he promotes inclusive, sustainable, and trustworthy AI infrastructure worldwide. 

\vspace{30pt}
\begin{wrapfigure}[11]{l}{0.23\textwidth}
    \vspace{-20pt}
    \includegraphics[width=1.5in,height=1.9in,clip,keepaspectratio]{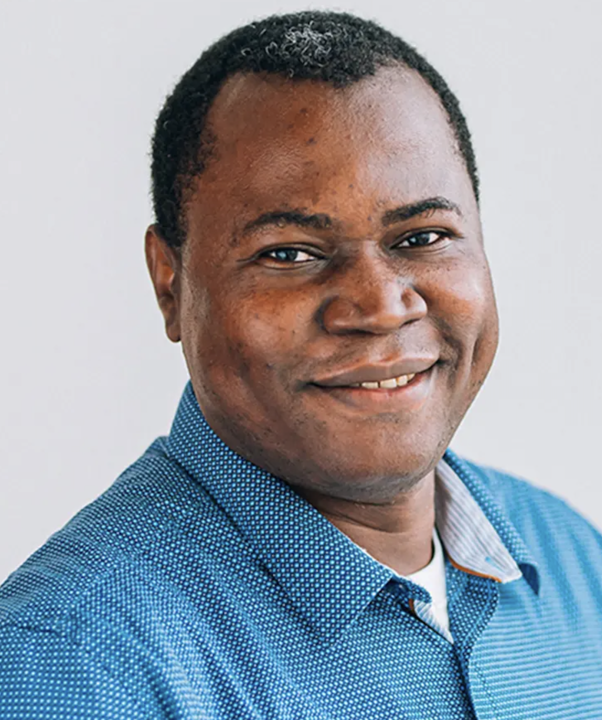}
    \vspace{-10pt}
\end{wrapfigure}
\textbf{Foutse Khomh} is a Full Professor of Software Engineering at Polytechnique Montréal, a Canada Research Chair Tier 1 on Trustworthy Intelligent Software Systems, a Canada CIFAR AI Chair on Trustworthy Machine Learning Software Systems, an NSERC Arthur B. McDonald Fellow, an Honoris Genius Prize Laureate, and an FRQ-IVADO Research Chair on Software Quality Assurance for Machine Learning Applications. He received a Ph.D. in Software Engineering from the University of Montreal in 2011, with the Award of Excellence. He also received a CS-Can/Info-Can Outstanding Young Computer Science Researcher Prize for 2019, the Excellence in Research and Innovation Award of Polytechnique Montréal, and the prestigious IEEE CS TCSE New Directions Award in 2025. His work has received four ten-year Most Influential Paper (MIP) Awards, eight Best/Distinguished Paper Awards at major conferences, and two Best Journal Paper of the Year Awards. He initiated and co-organized the Software Engineering for Machine Learning Applications (SEMLA) symposium and the RELENG (Release Engineering) workshop series. He also co-organized the FM+SE Summit series (https://fmse.io/), a platform where leading industrial and academic experts discuss and reflect on the challenges associated with the adoption of foundation and large models in software engineering. He is co-founder of the NSERC CREATE SE4AI: A Training Program on the Development, Deployment, and Servicing of Artificial Intelligence-based Software Systems and one of the Principal Investigators of the DEpendable Explainable Learning (DEEL) project. He is also a co-founder of Quebec's initiative on Trustworthy AI (Confiance IA Quebec) and Scientific co-director of the Institut de Valorisation des Données (IVADO). He is on the editorial board of multiple international software engineering journals (e.g., TOSEM, IEEE Software, EMSE, SQJ, JSEP) and is a Senior Member of IEEE.

\vspace{10pt}
\begin{wrapfigure}{l}{0.23\textwidth}
    \vspace{-20pt}
    \includegraphics[width=1.5in,height=1.9in,clip,keepaspectratio]{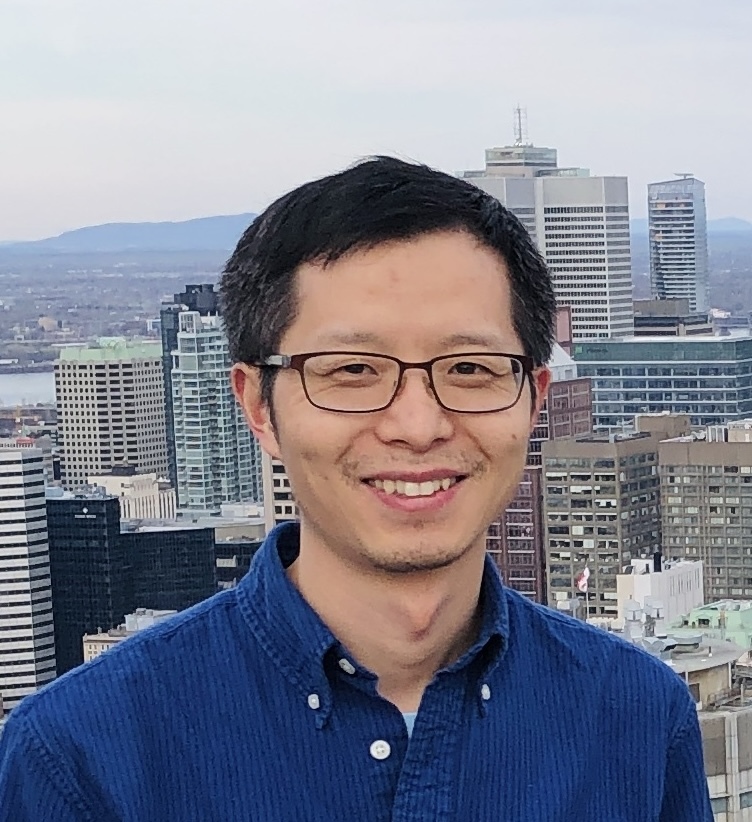}
\end{wrapfigure}
\textbf{Heng Li} is an associate professor in the Department of Computer and Software Engineering at Polytechnique Montreal. He holds a Ph.D. in Computing from Queen’s University (Canada), an M.Sc. from Fudan University (China), and a B.Eng. from Sun Yat-sen University (China). Before his academic career, he worked in the industry for years as a software engineer at Synopsys and a software performance engineer at BlackBerry. He and his students’ research in the MOOSE lab (moose.polymtl.ca/) aims to enhance the performance and reliability of modern software systems, primarily through performance engineering, runtime monitoring, and AI-powered analysis (AIOps).

\end{document}